\documentclass[amstex,epsf,amssymb,usenatbib,letterpaper]{aa}
\usepackage{epsf,graphicx}
\usepackage{amsmath,amssymb} 
\usepackage{subfigure}
\usepackage{natbib}
\bibpunct{(}{)}{;}{a}{}{,}

\usepackage{txfonts}

\begin{document}

   \title{The near-infrared excitation of the HH\,211 protostellar outflow 
   \thanks{Based in part on observations collected at the German-Spanish Astronomical 
   Center, Calar Alto, operated jointly by Max-Planck Institut f\"ur Astronomie and Instituto 
   de Astrof{\'\i}sica de Andaluc{\'\i}a (CSIC). Observations were also obtained with The United Kingdom 
   Infrared Telescope, which is operated by the Joint Astronomy Centre on behalf of the U.K. 
   Particle Physics and Astronomy Research 
   Council.}}

   \author{Barry O'Connell\inst{1,2} \and Michael D. Smith\inst{1} \and Dirk Froebrich\inst{3} \and 
   Christopher J. Davis\inst{4} \and Jochen Eisl{\"o}ffel\inst{5}}

   \institute{Armagh Observatory, College Hill, Armagh BT61 9DG, Northern
   Ireland, UK
   \and Physics Department, Trinity College Dublin, College Green, Dublin 2, Ireland
   \and School of Cosmic Physics, Dublin Institute for Advanced Studies, 
   5 Merrion Square, Dublin 2, Ireland
   \and Joint Astronomy Centre, 660 N.A'ohoku Place, University Park,
   Hilo, Hawaii 96720, USA 
   \and Th{\"u}ringer Landessternwarte Tautenburg, Sternwarte 5, 07778 Tautenberg, Germany 
    }

   \offprints{B.O'Connell, \email{boc@arm.ac.uk}}

   \date{Received ...... / Accepted ........}

   \abstract{The protostellar outflow HH\,211 is of considerable interest since it is
extremely young and highly collimated. Here, we explore the outflow through imaging and
spectroscopy in the near-infrared to determine if there are further diagnostic signatures of youth. 
We confirm the detection of a near-infrared continuum of unknown origin. We propose that it is
emitted by the driving millimeter source, escapes the core through tunnels, and 
illuminates features aligning the outflow. Narrow-band flux measurements of these features contain an 
unusually large amount of continuum emission. [Fe{\small II}] emission at 1.644\,$\mu$m has been detected and is restricted 
to isolated condensations. However, the characteristics of vibrational 
excitation of molecular hydrogen resemble those of older molecular outflows. 
We attempt to model the ordered structure of the western outflow as a series of shocks, 
finding that bow shocks with J-type dissociative apices and C-type flanks are consistent. 
Moreover, essentially the same conditions are predicted for all three bows except for a systematic 
reduction in speed and density with distance from the driving source. 
We find increased K-band extinctions in the bright regions as high as 2.9 magnitudes and suggest 
that the bow shocks become visible where the outflow impacts on dense clumps of cloud material. 
We propose that the outflow is carved out by episodes of pulsating jets. The jets, driven by
central explosive events, are responsible for excavating a central tunnel through which radiation 
temporarily penetrates.

               }
   \authorrunning{O'Connell et al.}
   \titlerunning{The HH\,211 protostellar outflow}

   \maketitle

\section{Introduction}

The processes controlling the birth of individual stars evade direct analysis in the optical because they 
operate within highly obscuring cores within dense molecular clouds. 
Nevertheless, the end of the period of gestation is heralded by often spectacular 
ejections of material \citep{1996ARA&A..34..111B}. We may thus deduce information 
concerning the accretion process by indirect means. Bipolar jets from protostars appear 
to be ubiquitous and their characteristics suggest a close relationship to the accretion 
process from a disc into a forming star \citep{2003MNRAS.346..163F}. 
In addition, they have drastic dynamical and chemical effects on their environment. 
They excavate cavities by sweeping up the molecular material which then forms an outflow 
visible in emission from CO molecules. As well as a leading bow wave which ploughs
through the ambient medium, irregularities in the outflow give rise to shocks which propagate 
away from the protostar into the ambient molecular gas \citep{2003Ap&SS.287..217F}. 
These bow shocks excite atoms and molecules, inducing line emission which is often visible at optical 
wavelengths as Herbig-Haro (HH) objects \citep{2001ARA&A..39..403R,2002RMxAC..13....1B}.  However,
the near-infrared emission is particularly important for the youngest
protostars since even the protruding outflows may still be obscured in the optical regime
\citep{2000prpl.conf..815E}.
In fact,  near-infrared observations reveal remarkable molecular hydrogen flows which hold valuable 
information concerning the environment through which they propagate as well as the mechanism which 
launches the driving jets \citep{2002RMxAC..13...36D}.

HH\,211 is a bipolar molecular outflow which was discovered by \citet{1994ApJ...436L.189M}. It lies near 
the young stellar cluster IC~348~IR in the Perseus dark cloud complex at an estimated distance of 315~pc 
\citep{1998ApJ...497..736H}. The outflow is bilaterally symmetric and highly collimated with an aspect ratio 
of $\sim$~15:1. The total extent of the outflow is 106\arcsec which is 0.16\,pc at the adopted distance. 
A H$_2$ (1,0)~S(1) wide-field survey of the IC~348 cluster was carried out by \citet{2003ApJ...595..259E} 
covering a 6\arcmin.8 $\times$ 6\arcmin.8 region; no HH~211 outflow remnants were detected beyond the outer 
knots. This makes it one of the smaller outflows, which suggests that it may also be one of the youngest 
since the average length within an unbiased sample of Class O/I jets was found to be 0.6--0.8\,pc 
\citep{2003Ap&SS.287..149S}.

The conclusion that HH~211 is a jet driven outflow with a timescale of order 1000 years was derived from 
interferometric CO observations \citep{1999A&A...343..571G}, which confirms it as one of the 
youngest infrared outflows to be discovered. Although the outflow lobes are visible in both blue-shifted 
and red-shifted CO emission, a small inclination angle to the plane of the sky is suggested by (i) the 
lack of strong differential extinction in the H$_2$ brightness distribution, (ii) the high degree of
separation of the blue and red CO lobes and (iii) the relatively small radial components of
the SiO and H$_2$ flow speeds \citep{2001ApJ...555..139C,2003RMxAA..39...77S}. 

The central engine driving the outflow is HH~211-mm, a low-mass protostar with a bolometric luminosity of 
3.6~L$_{\sun}$ and bolometric temperature of 33\,K. It is surrounded by a $\sim$~0.8~M$_{\sun}$ dust 
condensation \citep{Froebrich2005}\footnote{http://www.dias.ie/protostars/}.  
Since 4.6 per cent of the bolometric luminosity is attributed to the submillimeter 
luminosity, L$_{smm}$, HH~211-mm is classified as a Class O type protostar.

A compact and collimated SiO jet extends in both flow directions out to a projected distance of 20\arcsec~ from 
the central source \citep{1997IAUS..182P..76C,2001ApJ...555..139C}. This is also another indication of the Class
O nature of the source \citep{2004ApJ...603..198G}. The clumpy nature of the observed SiO emission suggests a 
shock origin resulting from a time dependent jet velocity. \citet{2002A&A...395L..25N} observed HH~211 in 
SiO lines originating from high rotational energy levels and deduced a high jet density 
of n$_{H2}$ $\sim$ 2--5 $\times$ 10$^6$ cm$^{-3}$ and gas temperature $\geq$~250\,K. However, no 
SiO emission is detected beyond 20\arcsec where the H$_2$ (1,0)~S(1) emission is found 
suggesting that the conditions, such as shock velocity and pre-shock density, vary considerably along the flow direction. 

Imaging of [Fe{\small II}] emission in the H-band at 1.644$\mu$m has recently played a major role in our 
understanding of shocked outflows (see \citet{2000AJ....120.1449R} for a summary). Where [Fe{\small II}]
emission is observed it traces the fast ($>$~50~km~s$^{-1}$) and dissociative shocks. Combined with K-band 
imaging of H$_2$ rotational-vibrational lines, which trace less extreme shock conditions, important
information about outflow excitation may be gathered \citep{2004A&A...418..163K}.  
 
Previous studies employing bow-shock models have focused on larger, more evolved systems
\citep{2000A&A...359.1147E,2003MNRAS.339..524S,2004A&A...419..975O}. HH~211 provides us with the unique opportunity of studying
what has been deemed as an exceptionally young outflow covering a small spacial extent.
We present here new high resolution images of H$_2$ and [Fe{\small II}] lines and K-band spectroscopy of the 
outflow (\S~3). We analyse the impact regions visible in the near-infrared. We then present in \S~4 C-type 
bow-shock models in 
order to interpret the remarkable set of bows propagating through a changing environment along the western 
outflow. In \S~5 we discuss the issues which have arisen from this set of data before pooling our findings
together in conjunction with previous studies to suggest a global outflow mechanism for
HH~211 in \S~6.        
 	     
\section{Observations and data reduction}
\label{observations}

\subsection{KSPEC observations}

\begin{figure*}[t]
\centering
\includegraphics[width=16.0cm]{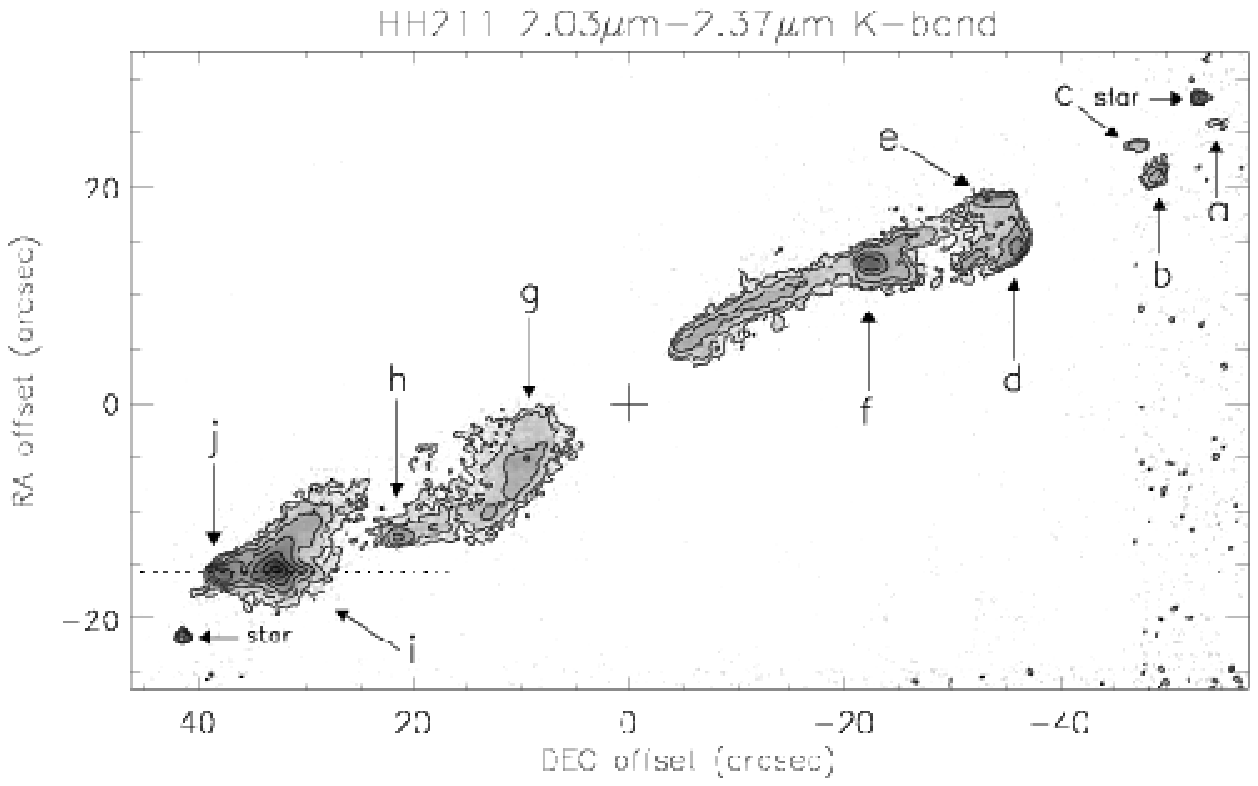}
\caption{Broad-band K image of HH\,211 which covers  wavelengths between 2.03$\mu$m and 2.37$\mu$m.
Features are labeled according to the nomenclature of \citet{1994ApJ...436L.189M}. The exciting source 
HH\,211-mm at position R.A.(2000) = 03$^h$ 43$^m$ 56.7$^s$, Dec(2000) = +32$\degr$ 00$\arcmin$ 50.3$\arcsec$ 
\citep{2001RMxAA..37..201A} is indicated by the cross. KSPEC spectroscopic slit positions are indicated by the dotted lines. 
Note that the star in the south-west of this image (and following images) has been masked for display purposes.}
\label{kbandimage}          
\end{figure*}

Our near-infrared (NIR) spectra cover the 1-2.5\,$\mu$m region in medium resolution. They were obtained in the period 
26-29 August 1996 with the KSPEC spectrograph on the University of Hawaii 2.2 meter telescope. This cross 
dispersed Echelle spectrograph is equipped with a HAWAII 1024\,$\times$\,1024 detector array and optimised 
for 2.2\,$\mu$m. Observations were performed at two bright H$_2$ emission locations. The  0.\arcsec96 width 
slits ran in an east-west direction passing through knots {\em f} and {\em d} in the west and through knots 
{\em i} and {\em j} in the east (positions are indicated in Fig.~\ref{kbandimage}).
Data reduction (flatfielding, sky-subtraction, extraction of the spectra) was performed using our own MIDAS 
routines. An absolute calibration of the fluxes was not possible due to non-photometric weather conditions.
Wavelength calibration was performed using OH-night-sky emission lines and tables of 
\citet{2000A&A...354.1134R}. The results are presented in Table~\ref{kspecfluxes}.  Note that fainter emission 
lines are detected in knots {\em i} and {\em j} due to the stronger emission here compared to knot 
{\em d} (see
Table \ref{tab:knots}.)  

\begin{table}[!h]
\caption{KSPEC relative fluxes. For both slit positions, the measured fluxes lie above the continuum
and are presented relative to the H$_2$ (1,0)~S(1) line flux. Three observations were carried out at 
each slit location. The relative fluxes have been averaged and the spread in values is quoted in brackets
as an error estimate.}
\begin{tabular}{c@{~~~~~~~~~}c@{~~~~~~~~~~~~~~}c@{~~~~~~~~~~~~~~}c}
\noalign{\smallskip}
\hline
\hline
\noalign{\smallskip}
Line &  $\lambda$($\mu$m) & east & west \\
\noalign{\smallskip}
\hline
\noalign{\smallskip}
(1,0)~S(9) & 1.687 & 0.02 (0.01) & -- \\
(1,0)~S(7) & 1.748 & 0.12 (0.01) & -- \\
(1,0)~S(6) & 1.788 & 0.07 (0.01) & -- \\
(1,0)~S(5) & 1.835 & 0.62 (0.04) & -- \\
(1,0)~S(4) & 1.891 & 0.20 (0.01) & -- \\ 
(1,0)~S(2) & 2.033 & 0.36 (0.01) & 0.30 (0.05) \\
(3,2)~S(5) & 2.065 & 0.03 (0.01) & -- \\
(2,1)~S(3) & 2.073 & 0.10 (0.01) & 0.18 (0.07) \\
(1,0)~S(1) & 2.121 & 1.00 (0.05) & 1.00 (0.05) \\
(3,2)~S(4) & 2.127 & 0.02 (0.01) & -- \\
(2,1)~S(2) & 2.154 & 0.05 (0.01) & -- \\
(3,2)~S(3) & 2.201 & 0.03 (0.01) & -- \\
(1,0)~S(0) & 2.223 & 0.25 (0.01) & 0.31 (0.02) \\ 
(2,1)~S(1) & 2.247 & 0.13 (0.01) & 0.17 (0.06) \\
(2,1)~S(0) & 2.355 & 0.03 (0.01) & -- \\
(3,2)~S(1) & 2.386 & 0.03 (0.01) & -- \\
(1,0)~Q(1) & 2.406 & 0.96 (0.05) & 1.25 (0.17) \\
(1,0)~Q(2) & 2.413 & 0.38 (0.05) & 0.54 (0.11) \\
(1,0)~Q(3) & 2.423 & 0.95 (0.05) & 1.13 (0.05) \\
(1,0)~Q(4) & 2.437 & 0.33 (0.04) & 0.41 (0.15) \\
(1,0)~Q(6) & 2.475 & 0.19 (0.06) & -- \\ 
(1,0)~Q(7) & 2.499 & 0.43 (0.03) & -- \\
\noalign{\smallskip}		       
\hline 
\end{tabular}
\label{kspecfluxes}
\end{table}

\subsection{MAGIC observations}

\begin{figure*}
\centering
\includegraphics[width=16cm]{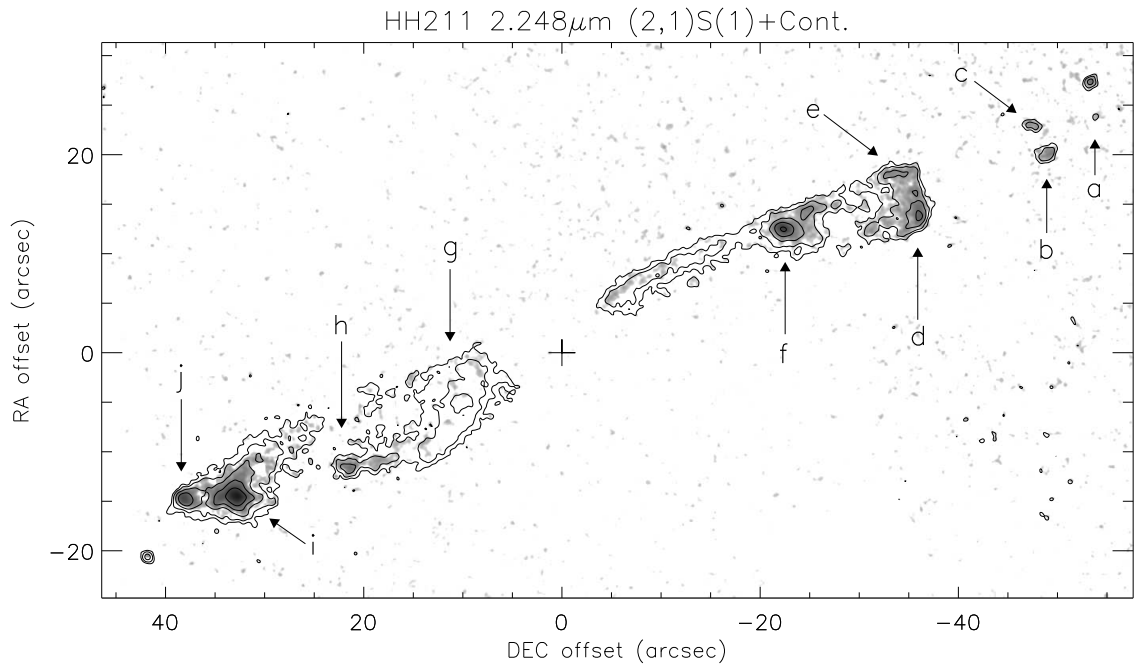}
\caption{HH\,211 at 2.248\,$\mu$m. The grey-scale is the continuum subtracted image showing only the 
(2,1)~S(1) line emission. Overlaid are contours representing the non continuum subtracted image. Thus 
the contours nearer the source trace predominantly scattered light while the contours further out mostly 
trace compact, line-emission features. The contour levels, which are scaled logarithmically, are 
at 0.22, 0.44, 0.87, 1.74, 3.47 $\times$ 10$^{-18}$ W m$^{-2}$ arcsec$^{-2}$. }
\label{twooneimage}
\end{figure*}

\begin{figure*}
\centering
\includegraphics[width=16cm]{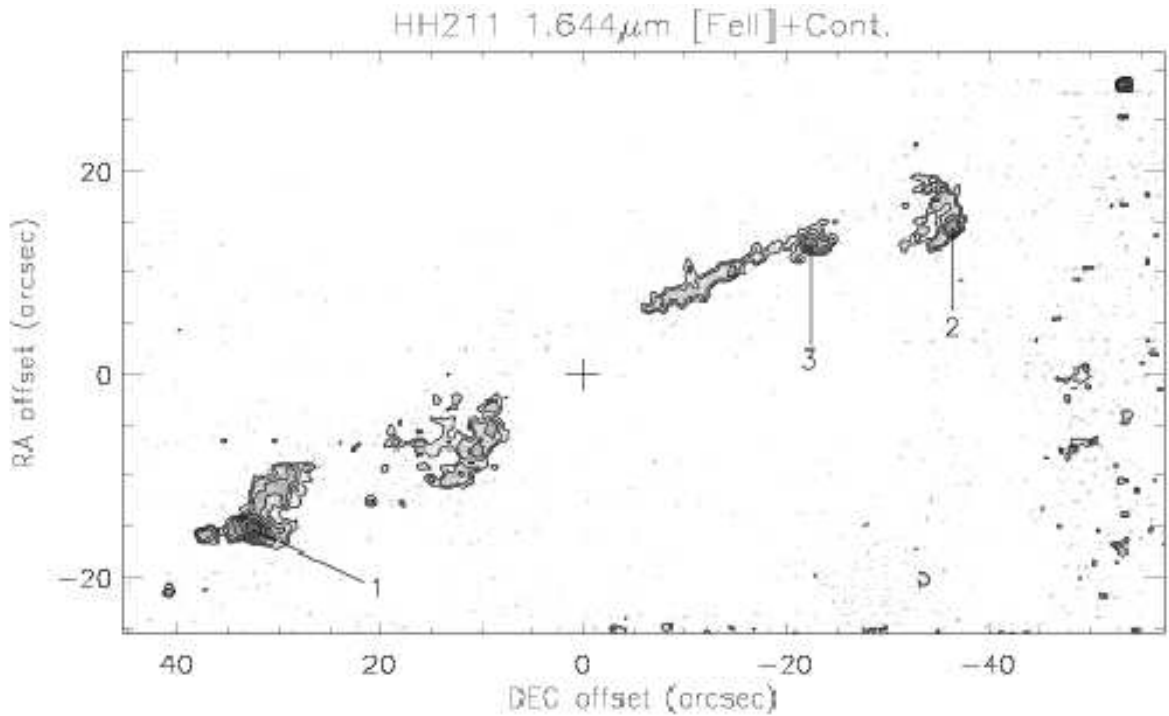}
\caption{HH\,211 at 1.644 $\mu$m. The greyscale {\em and} contours both trace the [Fe{\small II}] line plus continuum
emission. Consequently, most of the distributed flux can be attributed to continuum emission, although some concentrated
condensations of [Fe{\small II}] line emission are observed. These are labelled 1 --3. The logarithmic contours measure 
the flux at 0.63, 1.0, 1.58, 2.51, 3.98, 
6.31 $\times$ 10$^{-19}$~W~m$^{-2}$~arcsec$^{-2}$.  }
\label{feimage}
\end{figure*}

The NIR images were taken in November 1995 at the 3.5 meter telescope on Calar Alto using the MAGIC 
infrared camera \citep{1993SPIE.1946..605H} in its high resolution mode (0.32\arcsec~ per pixel). 
Images were obtained using narrow band filters centered on the H$_2$ (1,0)~S(1) emission line at 
2.122\,$\mu$m, the (2,1)~S(1) line at 2.248\,$\mu$m, the (3,2)~S(3) line at 2.201\,$\mu$m and on the 
nearby continuum at 2.140\,$\mu$m. The per pixel integration time was 1740s. Seeing throughout the 
observations was $\sim$0.9\arcsec except for the 2.14 $\mu$m image where it is $\sim$1\arcsec.8. The narrow-band 
image containing the (1,0)~S(1) line has been previously 
published in \citet{2003ApJ...595..259E}.

The data could not be accurately flux calibrated due to non-photometric conditions. The total H$_2$ 
(1,0)~S(1) flux from the entire outflow was previously measured to be 1.0\,$\times$\,10$^{-15}$ W~m$^{-2}$ 
by \citet{1994ApJ...436L.189M} which agrees with our flux calibration. We have deduced a calibration for 
all our narrow band images according to this measurement assuming that the average integrated flux from 
several bright unsaturated field of view stars should be similar for each filter. This should be the
case because each filter FWHM is equal to 0.02\,$\mu$m and the spectral energy
distribution (SED) is likely to be relatively flat on average between 2.122\,$\mu$m and 2.248\,$\mu$m.
We assume the accuracy of this method to be $\sim$15\%.

\subsection{UFTI observations}

Further NIR observations of HH\,211 were carried out on December 12, 2000 (UT) at the U.K. Infrared 
Telescope ({\em UKIRT}) using the near-infrared Fast Track Imager UFTI 
\citep{2003SPIE.4841..901R}. The camera is equipped with a Rockwell Hawaii 1024\,$\times$\,1024 
HgCdTe array which has a plate scale of 0\arcsec.091 per pixel and provides a total field of view of 
92\arcsec.9 $\times$ 92\arcsec.9.

Images in the [Fe{\small II}] ${^4}$D$_{7/2}$ -- ${^4}$F$_{9/2}$ transition were obtained
using a narrow-band filter centered on $\lambda$ = 1.644\,$\mu$m with $\Delta \lambda$(FWHM) =
0.016\,$\mu$m). The outflow was also imaged using the broad-band K[98] filter centered on 
$\lambda$ = 2.20\,$\mu$m with $\Delta\lambda$(FWHM) = 0.34$\mu$m. Seeing throughout the observations 
was $\sim$0\arcsec.8.~ Nine-point `jittered' mosaics were obtained in each filter.

Standard reduction techniques were employed (using the Starlink packages CCDPACK and KAPPA) including 
bad-pixel masking, sky subtraction and flat-field creation (from the jittered source frames themselves). 
The images were registered  using common stars in overlapping regions and mosaicked. The observations 
were conducted under photometric conditions, so the faint standard FS11 (spectral type A3; H = 11.267 mag) 
\citep{2001MNRAS.325..563H} was also observed and used to flux calibrate the [Fe{\small II}] image. 
The final images were smoothed using a circular Gaussian filter of FWHM = 0\arcsec.25.~ 
in order to increase the signal to noise without compromising the resolution.

\section{Results}
\label{results} 

\begin{table*}
\caption{Photometric results for HH\,211. The flux in units of 10$^{-18}$\,W\,m$^{-2}$ is measured over the 
indicated circular apertures. Note that the measurements were made from images which are not continuum 
subtracted because of the low S/N of the continuum image. To infer the line emission fluxes the continuum values$^{\ast}$ need to be subtracted. 
The (2,1)~S(1)/(1,0)~S(1) (labeled 2/1) ratios have been determined after subtracting the 2.14\,$\mu$m 
continuum emission. In the case of knot {\em g} the (2,1)~S(1) flux is equal to the continuum flux so 
no ratio was derived.} 
\begin{tabular}{c@{~~~~~}c@{~~~~~~}c@{~~~~~~}c@{~~~~~~}c@{~~~~~~}c@{~~~~~~}c@{~~~~~~}c@{~~~~~~}c@{~}}
\noalign{\smallskip}
\hline
\hline
\noalign{\smallskip}
Knot & aperture & (1,0)~S(1)$^{\dagger}$ & (2,1)~S(1)$^{\dagger}$ & (3,2)~S(3)$^{\dagger}$ & 2.14$\mu$m 
Cont.$^{\dagger\ast}$ & Broad-band K$^{\ddagger}$ & [FeII] 1.644$\mu$m$^{\dagger}$ & 2/1 ratio\\
\noalign{\smallskip}
\hline
\noalign{\smallskip}
HH~211-west & 57\arcsec & 413 & 61 & 34 & 42 & 1285 & 28 & 0.05~(0.04)\\
HH~211-east & 42\arcsec & 575  & 98  & 63 & 55 & 2197& 41 & 0.08~(0.03)\\
bow-{\em a} & 5.3\arcsec & 7.2 & 0.6 & no det. & no det. & 9 & no det. & 0.08~(0.04)\\
bow-{\em bc} & 7.7\arcsec & 32.0  & 2.3 & no det. & no det. & 65 & no det. & 0.07~(0.02)\\
bow-{\em de} & 10.1\arcsec & 155.5 & 18.1 & 7.3 & 3.1 & 335 & 11.8 & 0.09~(0.02) \\
{\em b} & 4.9\arcsec & 19.3  & 1.5 & no det. & no det. & 41 & no det.  & 0.08~(0.02) \\
{\em c} & 4.9\arcsec & 11.2  & 0.9 & no det. & no det. & 21 & no det.  & 0.08~(0.03) \\
{\em d} & 6.0\arcsec & 89.6  & 10.2 & 4.0 & 1.9 & 192 & 6.1  & 0.09~(0.03) \\
{\em e} & 6.0\arcsec & 57.5  & 7.9 & 3.3 & 1.3 & 131 & 4.8  & 0.12~(0.03) \\
{\em f} & 10.1\arcsec & 157.2  & 19.3 & 9.1 & 8.9 & 395 & 9.0  & 0.07~(0.04) \\
{\em g} & 10.1\arcsec & 41.9  & 20.6 & 20.6 & 19.6 & 380 & 14.3 & -- \\
{\em h} & 10.1\arcsec & 60.3  & 9.8 & 5.1 & 4.7 & 190 & 4.6 & 0.09~(0.08) \\
{\em ij} & 17.9\arcsec & 458.3 & 54.3 & 27.9 & 24.6 & 1364 & 25.0 & 0.07~(0.03) \\ 
\noalign{\smallskip}
\hline
\noalign{\smallskip}
\end{tabular}
\footnotesize $^{\dagger}$ The background 1$\sigma$ noise estimates over an 8\arcsec aperture are: 2.8
for H$_2$ (1,0)~S(1); 0.2 for H$_2$ (2,1)~S(1);
0.7 for H$_2$ (3,2)~S(3); 1.1 for the 2.14$\mu$m 
continuum;
and 3.4 for the [Fe{\small II}] at 1.644$\mu$m, also in units of 10$^{-18}$\,W\,m$^{-2}$.\\
\footnotesize $^{\ddagger}$ As a standard star was not observed in the K[98] filter the broad-band fluxes are rough
estimates. The flux calibration factor which we used was derived from the (1,0)~S(1) calibration
and a comparison of the broad-band and narrow-band filters. 
\label{tab:knots}
\end{table*}

Fig.~\ref{kbandimage} displays the HH~211 outflow in the K-band between 2.03$\mu$m and 2.37$\mu$m which
contains all the K-band line emission as well as a large proportion of continuum emission. The principle 
knots have been labeled as in \citet{1994ApJ...436L.189M}.

The H$_2$ (2,1)~S(1) image is displayed in Fig.~\ref{twooneimage}. The continuum at 2.14\,$\mu$m has 
been subtracted in order to indicate locations of pure (2,1)~S(1) emission (greyscale). Contours of 
the non continuum subtracted image are also displayed in order to indicate the extent of the continuum 
emission at 2.248\,$\mu$m.  The line emission is produced from an excitation level which is 12,553\,K 
above the ground state whereas the (1,0)~S(1) arises from 6,953\,K. Therefore, we expect it to highlight 
the hotter parts of molecular shocks.

The western outflow shows particularly interesting structures which can be described as a series of bow 
shocks propagating along the outflow away from the source. Bows {\em de} and {\em bc} display a common 
asymmetric structure: the lower bow wing is approximately 1.5 times brighter than the upper wing. 
We will interpret this asymmetry in \S~\ref{analysis} as due to a misalignment of the magnetic field with 
the flow through which the bow shock configurations with C-type flanks are propagating. 

Emission detected at 1.644\,$\mu$m is displayed in Fig.~\ref{feimage}. Most of the emission detected here is 
attributed to the high level of continuum flux, as is seen in the K-band (Fig.~\ref{kbandimage}). Steeply 
rising above this continuum level are several concentrated [Fe{\small II}] emission condensations, labeled 1 -- 3, 
one of which forms part of a well defined bow-shock, bow-{\em de}.

We confirm the existence of a band of continuum emission which extends along the western 
outflow. It becomes visible 5\arcsec~ from the driving source and maintains a relatively constant flux 
out to 17\arcsec~ from the source. \citet{2003ApJ...595..259E} suggest that this continuum is scattered
radiation from HH\,211-mm which opens the possibility of indirectly obtaining a spectrum of the outflow source.
The band of continuum is prominent in the K-band as well as at 1.644$\mu$m.

The photometric results for knots {\em a} -- {\em j} are listed in Table~\ref{tab:knots}. Knots {\em d,e} and 
{\em b,c} are interpreted as bow-shock components and are labeled as bow-{\em de} and bow-{\em bc}. 
We will briefly discuss the implications of these fluxes. The eastern outflow is $\sim$1.5 times brighter than
the western outflow. However, the cause of this difference is not necessarily due to an unequal jet power output as 
\citet{2001ApJ...555...40G} detected similar levels of {O\small I} 63\,$\mu$m in both lobes (1.02\,L$_{\sun}$ in east; 
0.94\,L$_{\sun}$ in west). This line is relatively unaffected by extinction and represents the main cooling channel in the 
post-shocked gas. The non photometric conditions during the observations are reflected in the relatively large
1$\sigma$ noise levels which have been estimated from the background flux variation using an 8\arcsec aperture. Most of the 
emission in the (3,2)~S(3) image is actually continuum emission except possibly for bow-{\em de}. Interestingly, the fluxes at 
1.644\,$\mu$m are also comparable to the continuum fluxes. This implies that the dereddened continuum fluxes are 
2 -- 5 times brighter at 1.644\,$\mu$m than at 2.14$\mu$m taking into consideration the higher extinction (A$_H$ =
1.6 A$_K$), the range of extinctions explored (see below and \S~\ref{Discussion}), and that the 1.644$\mu$m 
filter width is 25\% narrower than the 2.14$\mu$m filter. The (2,1)~S(1)/(1,0)~S(1) ratios are calculated with 
the continuum subtracted. The error propagation results in very large errors at locations where the continuum 
forms a large fraction of the H$_2$ emission.        

From the KSPEC data, we can extract several pieces of information. 
We calculate the differential extinction between two transition lines (in magnitudes) using
\begin{equation}
   \Delta=2.512 \, \log \left( \frac{F_1 \lambda_1 g_2 A_2}{F_2 \lambda_2 g_1 A_1} \right)
\end{equation} 
where $\lambda_1$ and $\lambda_2$ are the transition wavelengths, F$_1$ and F$_2$ are the relative 
fluxes,  g$_1$ and g$_2$ are the upper level degeneracies, and A$_1$ and A$_2$ are the spontaneous 
electric quadrupole transition probabilities taken from 
\citet{1998ApJS..115..293W}.

When both lines originate from the same upper level and a K-band spectral index of 1.7 is adopted, the 
absolute extinction is given by
\begin{equation}
    A_K = \frac{\Delta}{\left[ \left( \frac{2.2\mu m}{\lambda_2} \right)^{1.7} -   
          \left( \frac{2.2\mu m}{\lambda_1} \right)^{1.7} \right]}.
\end{equation}
Three pairs of H$_2$ $v$=(1,0) lines from the KSPEC data were used in order to determine 
A$_K$: Q(2)/S(0), Q(3)/S(1) and Q(4)/S(2). The results are presented in Table~\ref{Akvalues}.

\begin{table}[!h]
\caption{A$_K$ values (in magnitudes) which have been determined from the KSPEC relative fluxes in 
Table~\ref{kspecfluxes}. The flux errors have been propagated and yield a realistic A$_K$ error estimate.}
\begin{tabular}{c@{~~~~~~~~~}c@{~~~~~~~~~~~~~~}c@{~~~~~~~~~~~~~~}c}
\noalign{\smallskip}
\hline
\hline
\noalign{\smallskip}
& Q(2)/S(0) &  Q(3)/S(1) & Q(4)/S(2)  \\
\noalign{\smallskip}
\hline
\noalign{\smallskip}
A$_K$ east & 2.7 $\pm$ 1.5 & 1.5 $\pm$ 0.5 & 1.8 $\pm$ 0.5 \\
\noalign{\smallskip}
A$_K$ west & 3.9 $\pm$ 2.3 & 2.4 $\pm$ 0.5 & 3.2 $\pm$ 2.0 \\
\noalign{\smallskip}		       
\hline 
\end{tabular}
\label{Akvalues}
\end{table} 

Immediately evident is that the extinction is higher in the western knots although the extinction 
determination in the west is rather tentative due to the observing conditions. A higher extinction 
than in the east is plausible since the western lobe is redshifted and the embedded cloud (observed 
in H$^{13}$CO$^{+}$ by \citet{1999A&A...343..571G}) extends predominantly in this direction.
In order to determine the average value of A$_K$ each line pair was assigned with the relative line 
strength as a statistical weight. The statistically weighted average A$_K$ values for the eastern and 
western components are 1.8 and 2.9 magnitudes respectively. These values fall within the error limits 
and are used to deredden luminosities in \S~\ref{analysis}.  

\begin{figure}
   \includegraphics[width=8.5cm]{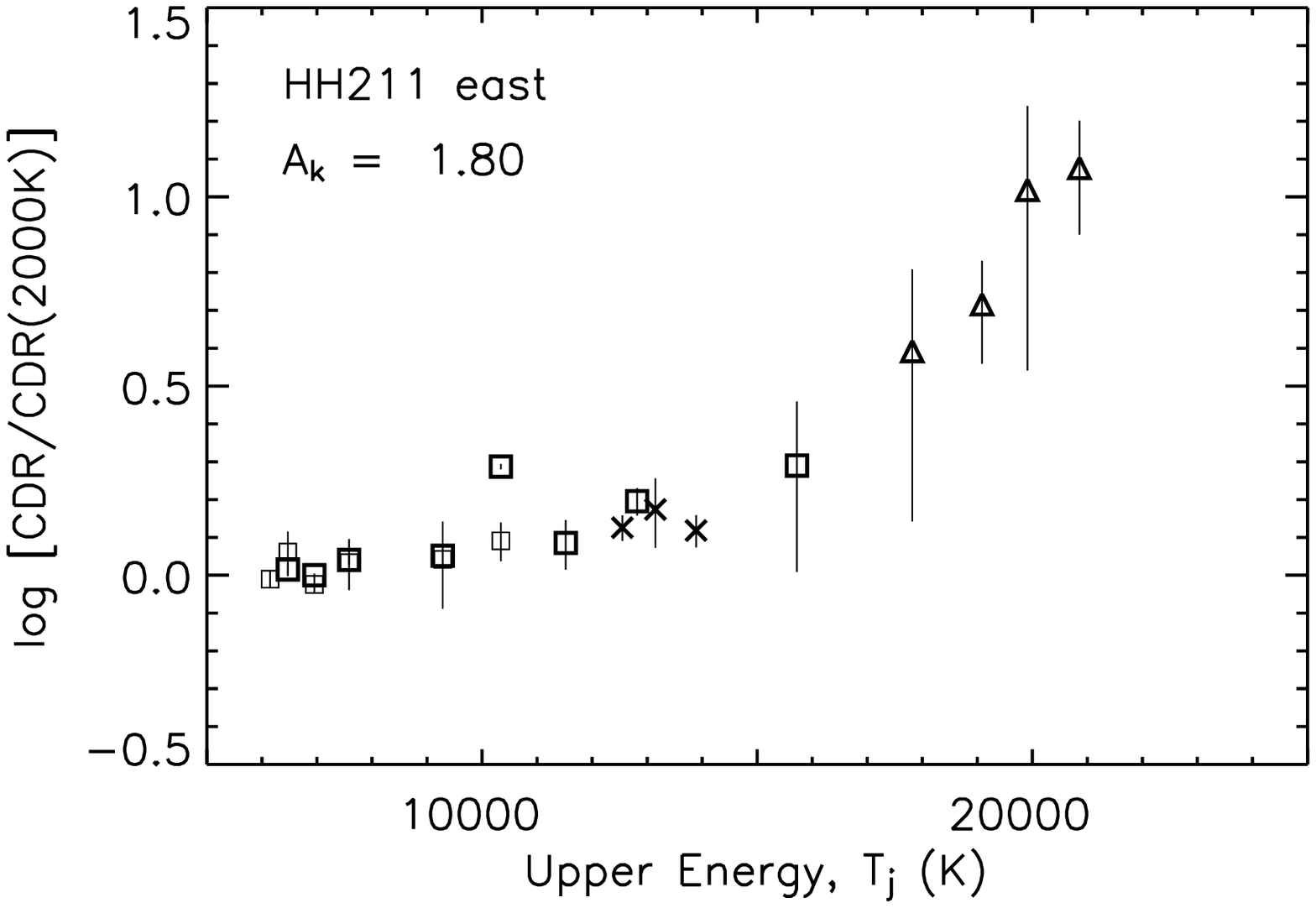}
   \includegraphics[width=8.5cm]{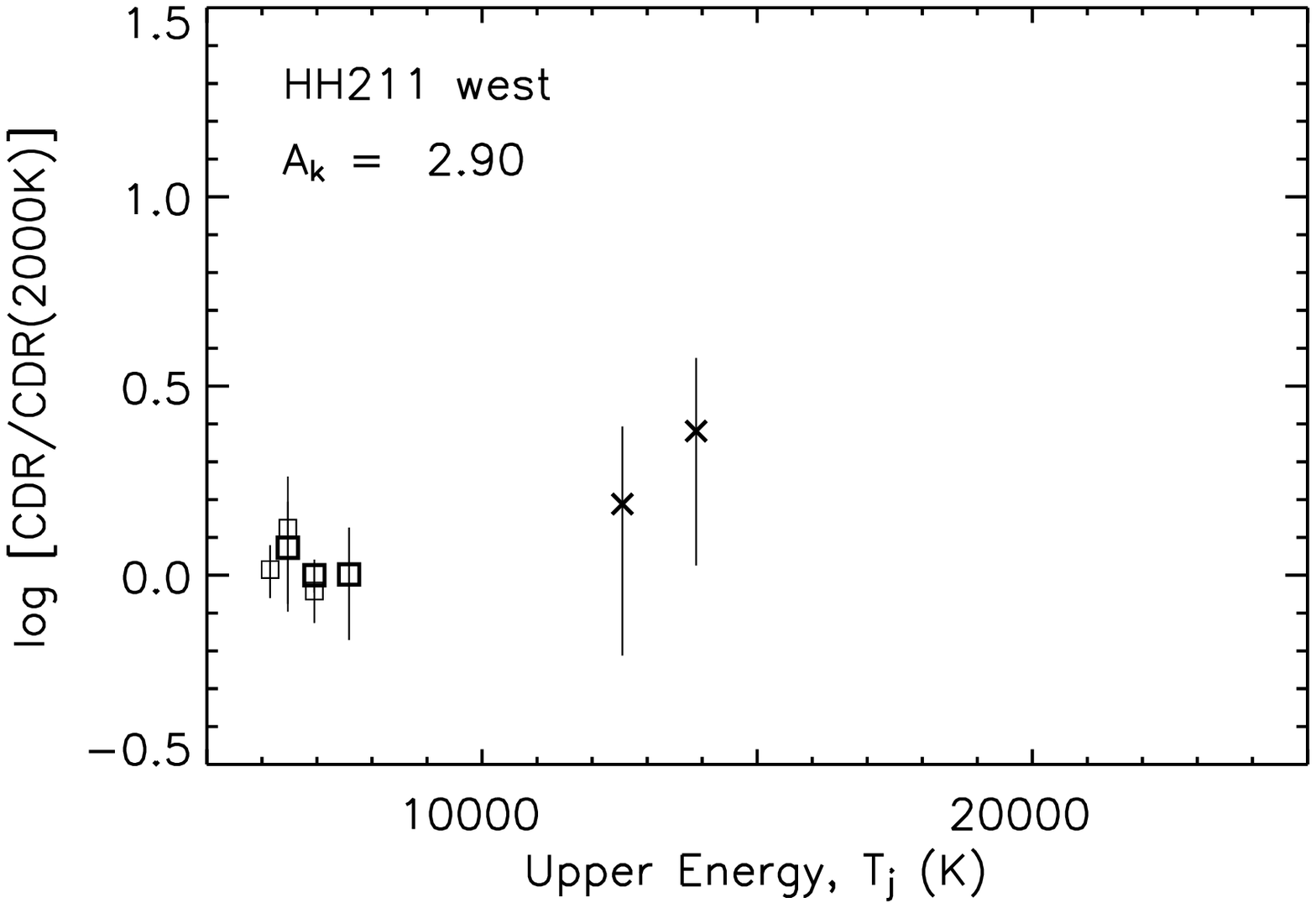}
      \caption{Column density ratio diagrams for HH\,211, produced from vertical slits running
through locations of peak emission. The extinction has been adjusted
to minimise the differences in 1--0 S-branch and Q-branch lines originating from the same upper
energy level. H$_2$ (1,0) transitions are represented by squares, (2,1) transitions by crosses, and (3,2), (3,1) and (3,0)
transitions by triangles. The faint squares represent (1,0) Q branch measurements. }
\label{cdrdiags}       
\end{figure}

In order to interpret the data, we employ the Column Density Ratio (CDR) method. We determine the column
of gas, $N_j$, in the upper energy level, $k\,T_j$, necessary to produce each line. We then
divide these values by the columns predicted from a gas at 2,000\,K in local thermodynamic equilibrium
with an ortho to para ratio of 3. Normalising to the (1,0)~S(1) line, we then plot this as the CDR,
as displayed in Fig.~\ref{cdrdiags}. We immediately see that the CDR is not constant but a
function of excitation temperature, and thus is not consistent with emission from a
uniform temperature region or from a single planar shock. Furthermore, no significant
deviations from a  single curve are identified (apart from that derived from the (1,0)~S(5)
which is not confirmed by the (1,0) Q(7) from the same $T_j$). Hence, the ortho to para ratio
is consistent with the value of three, usually associated with H$_2$ shocks.  

The extinction determined in the east location also constrains the H-band lines. A 
significantly lower extinction would raise the CDRs for these (1,0) H$_2$ above that of the (2,1)
K-band lines which would have implied non-LTE low density conditions. As it stands, the fact that
a single curve is predicted suggests a density sufficiently high to ensure LTE. Given a high fraction of 
hydrogen atoms, the lower vibrational levels of hydrogen molecules reach LTE at densities above
$\sim$~10$^{4}$~cm$^{-3}$.

\section{Analysis}
\label{analysis}

\subsection{Modeling the bow shocks}

\begin{figure}
\includegraphics[width=9cm]{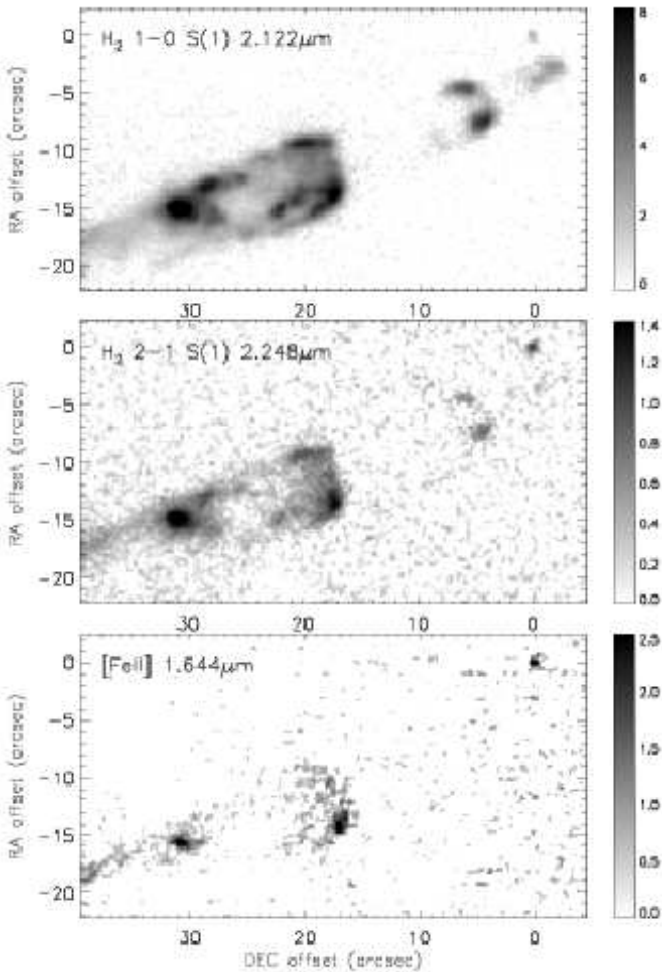}
\caption{The HH\,211 western outflow is shown here in H$_2$ (1,0)~S(1), H$_2$ (2,1)~S(1) and [Fe{\small II}] 
emission lines. The images are not continuum subtracted as the bow shocks are relatively pure. Greyscale bars 
represent flux levels per arcsec$^{-2}$ ~in units of 10$^{-18}$ W~m$^{-2}$.}
\label{western}
\end{figure}

The structure of the western outflow, shown in detail in Fig.~\ref{western}, raises an ideal 
interpretation scenario. The excited H$_2$ can be found in several distinct knots along an outflow axis. 
\citet{1994ApJ...436L.189M} suggested that this well organised
appearance might prove particularly amenable to modeling. A series of bow-shocks is propagating 
along the outflow. They gradually exhaust their momentum and slow down as they plough through
either ambient gas or the material in the wakes of upstream bow shocks. They encounter 
less dense gas towards the edge of the cloud. The submillimeter observations of 
\citet{2000ApJ...530..851C} show that the HH~211-mm envelope density decreases with distance
from the source. The azimuthally averaged density structure is well fitted by a single power-law, \, 
$\rho \propto r^{-1.5}$, out to 0.1~pc (the projected distance of the outer knots of the outflow from the 
source). However, the density profile is extended in a direction perpendicular to the outflow and the 
outer western knots are located outside this dense core region. 

The bow-shock models described by \citet{2003MNRAS.339..524S} have been successfully employed to 
interpret the sequence of bow shaped features along the HH\,240 outflow \citep{2004A&A...419..975O}. 
Here we apply the same model which serves as a useful 
interpretive device. We wish to determine if a systematic change of one or more parameters results in 
a close match with the observed bow structures and thus to analyse the outflow's changing environment. We have 
applied the bow-shock model to the three leading structures: {\em de}, {\em bc} and {\em a}. 

Jump-type (J-type) bow shocks can explain many of the features associated with high excitation emission regions.
They cause rapid heating and dissociate H$_2$ molecules for shock speeds greater than 24 km s$^{-1}$ 
\citep{1977ApJ...216..713K}. However, Continuous-type (C-type) bow shocks have proved extremely successful at 
interpreting most of the observed structures. 
The measured low fraction of ions in molecular clouds is consistent with their application. The magnetic field 
cushioning means that less energy goes into molecule dissociation; they can explain the high infrared fluxes
which are observed in bow shocks. However, their observed velocities often exceed the H$_2$ dissociation speed of 
$\sim$ 40--50 km s$^{-1}$ \citep{1991MNRAS.248..451S} giving rise to a double-zone bow shock composed of (1) a curved 
J-type dissociative cap (responsible for atomic emission) and (2) C-type wings (where H$_2$ emission is radiated).   

Our C-type bow shock model consists of a three dimensional curved surface described by 
\begin{equation}
   Z/L_{bow} = (1/s)(R/L_{bow})^{s},
\end{equation} 
where $Z$ and $R$ are cylindrical coordinates, $L_{bow}$ characterises the bow size  and $s$ defines the 
shape or sharpness of the bow. The curved shock surface is then divided into a very large number of
steady-state planar shocks for which the detailed physics and chemistry are computed 
\citep{1980ApJ...241.1021D,1991MNRAS.248..451S}. Each mini-shock propagates at a different velocity depending 
on the angle between the shock surface normal and the direction of motion of the bow. The temperature reached, 
and hence the excitation conditions, depends on the shock velocity, density, magnetic field strength, etc., as 
well as the magnetic field direction (see Table~\ref{cmodels}).



\begin{table}
    \caption{Model parameters derived to fit the bow images with C-type shocks.}
\label{cmodels}
   \begin{tabular}{llll}
\noalign{\smallskip}   
\hline
\hline
\noalign{\smallskip} 
Parameter & bow-{\em de} & bow-{\em bc} & bow-{\em a} \\
\noalign{\smallskip} 
\hline
\noalign{\smallskip} 
Size, $L_{bow}$ (cm)   & 1.0 $\times$ 10$^{16}$  & 1.0 $\times$ 10$^{16}$  &   1.0 $\times$ 10$^{16}$  \\
H Density, n (cm$^{-3}$)  & 8.0 $\times$ 10$^{3}$\, & 4.0 $\times$ 10$^{3}$  & 3.0 $\times$ 10$^{3}$ \\
Molecular Fraction      & 0.2              & 0.2               &   0.2  \\
Alfv\'en Speed, $v_A$ (km\,s$^{-1}$)  & 4      & 4     &  4      \\
Magnetic Field ($\mu$G)    & 193     & 137      &   118   \\
Ion Fraction, $\chi$    & 1.0 $\times$ 10$^{-5}$    & 3.0 $\times$ 10$^{-5}$  & 5.5 $\times$ 10$^{-5}$   \\
Bow Velocity, $v_{bow}$ (km\,s$^{-1}$) & 55      & 40           &   29   \\
Angle to l.o.s.          & 100$^{\circ}$            & 100$^{\circ}$            &   100$^{\circ}$   \\
s Parameter             & 2.10                & 1.90                 &   1.75  \\
Field angle, $\mu$      & 60$^{\circ}$        & 60$^{\circ}$         &   60$^{\circ}$  \\
\noalign{\smallskip}   
\hline  
\end{tabular}
\end{table}

\begin{table}
\caption{Observed and predicted bow shock luminosities and (2,1)~S(1) /
(1,0)~S(1) flux ratios. Luminosities are expressed in units of L$_{\sun}$ and a
distance of 315 pc is adopted. Luminosities have been dereddened using K-band
and H-band extinctions of 2.9 and 4.5 magnitudes (using A$_H$ = 1.56 $\times$ A$_K$ from 
\citet{1985ApJ...288..618R}) although the errors in these values are relatively large, see table~\ref{Akvalues}.
respectively. Note that the [Fe{\small 
II}] luminosities have been predicted using a J-type shock model.}
\label{lumtable} 
\begin{tabular}{c@{~~~~~~}c@{~~~~~~}c@{~~~~~}c@{~~}}
\noalign{\smallskip}
\hline
\hline
\noalign{\smallskip}
Line           & Observed$^{\ast}$    & Dereddened           & C-type  \\
               &                      &                      & Model   \\
\noalign{\smallskip}
\hline
\noalign{\smallskip}
 bow-{\em de}         &                      &                      & \\
\noalign{\smallskip}
\hline
\noalign{\smallskip}
H${_2}$~(1,0)~S(1) & 4.6 $\times$ 10$^{-4}$ & 6.5 $\times$ 10$^{-3}$ & 6.8 $\times$ 10$^{-3}$   \\
H${_2}$~(2,1)~S(1) & 4.3 $\times$ 10$^{-5}$ & 6.1 $\times$ 10$^{-4}$ & 7.3 $\times$ 10$^{-4}$    \\
H${_2}$~(3,2)~S(3) & 9.8 $\times$ 10$^{-6}$ & 1.4 $\times$ 10$^{-4}$ & 1.6 $\times$ 10$^{-4}$    \\
$[FeII]$ ~ ${^4}$D$_{7/2}$ -- ${^4}$F$_{9/2}$ & 3.6 $\times$ 10$^{-5}$ & 2.2 $\times$ 10$^{-3}$ & 
2.3 $\times$ 10$^{-3}$  \\
$\mathsf{2/1~~ ratio}$ & $\mathsf{0.09 \, (0.02)}$     & $\mathsf{0.09 \, (0.02)}$     & $\mathsf{0.11}$ \\

\noalign{\smallskip}	  
\hline
\noalign{\smallskip}
 bow-{\em bc}         &                      &                            &   \\
\noalign{\smallskip}		       
\hline
\noalign{\smallskip}
H${_2}$~(1,0)~S(1) & 9.7 $\times$ 10$^{-5}$ & 1.4 $\times$ 10$^{-3}$ & 1.4 $\times$ 10$^{-3}$   \\
H${_2}$~(2,1)~S(1) & 6.85 $\times$ 10$^{-6}$ & 9.7 $\times$ 10$^{-5}$ & 1.5 $\times$ 10$^{-4}$    \\
H${_2}$~(3,2)~S(3) & no det.$^{\ast}$ & -- & 2.7 $\times$ 10$^{-5}$    \\
$[FeII]$ ${^4}$D$_{7/2}$ -- ${^4}$F$_{9/2}$ & no det.$^{\ast}$ & -- & 1.2 $\times$ 10$^{-4}$   \\
$\mathsf{2/1~~ ratio}$ & $\mathsf{0.07 \, (0.02)}$     & $\mathsf{0.07 \, (0.02)}$     & $\mathsf{0.12}$ \\

\noalign{\smallskip}	  
\hline
\noalign{\smallskip}
bow-{\em a}         &                      &                            &   \\
\noalign{\smallskip}		       
\hline
\noalign{\smallskip}
H${_2}$~(1,0)~S(1) & 2.2 $\times$ 10$^{-5}$ & 3.1 $\times$ 10$^{-4}$ & 3.2 $\times$ 10$^{-4}$   \\
H${_2}$~(2,1)~S(1) & 1.7 $\times$ 10$^{-6}$ & 2.4 $\times$ 10$^{-5}$ & 2.5  $\times$ 10$^{-5}$   \\
H${_2}$~(3,2)~S(3) & no det.$^{\ast}$ & -- & 2.6 $\times$ 10$^{-6}$    \\
$[FeII]$ ${^4}$D$_{7/2}$ -- ${^4}$F$_{9/2}$ & no det.$^{\ast}$ & -- & 1.2 $\times$ 10$^{-6}$   \\
$\mathsf{2/1~~ ratio}$ & $\mathsf{0.08 \, (0.04)}$     & $\mathsf{0.08 \, (0.02)}$     & $\mathsf{0.08}$ \\
\noalign{\smallskip}	   
\hline 
\noalign{\smallskip}
\end{tabular}
\footnotesize $^{\ast}$The 3$\sigma$ detection limits for the observed knots over an 8\arcsec circular aperture
are: 2.6 $\times$ 10$^{-5}$ for (1,0)~S(1); 1.8 $\times$ 10$^{-6}$ for (2,1)~S(1); 6.4 $\times$ 10$^{-6}$ for
(3,2)~S(3); and 3.1 $\times$ 10$^{-5}$ for the [Fe{\small II}] image.
\end{table}

The  systematic method of exploration of parameter space is given in \citet{2004A&A...419..975O}.
In summary, the bow luminosities provide constraints on the density and bow speed. The location of
the emission in the flanks or apex also constrains the bow speed. In addition, the ion fraction 
constrains the transverse bow thickness as well as the bow speed. The magnetic field strength strongly
influences the extent of the wing emission and the atomic fraction affects the line ratios. As
demonstrated by previous modelling, the bow speed is limited to within $\sim$~15\%, the orientation to
within 10$^\circ$ and the density, magnetic field, atomic fraction and ion fraction to within a factor of two. 

The observed H${_2}$~(1,0)~S(1) luminosity for each bow provides us with the strongest density and 
velocity constraint. For a given bow size, L$_{bow}$, the line emission is directly proportional 
to the mass density (= 2.32 $\times$ 10$^{-24}$n$_c$) and ($v_{bow}$)$^3$. The observed 
bow luminosities have been dereddened using a K-band extinction of A$_K$~=~2.9
mag (see \S~\ref{results}). Table~\ref{lumtable} lists the observed dereddened 
luminosities for each knot along with the predicted model luminosities.

The parameters selected in order to model each bow are given in Table \ref{cmodels}. A constant Alfv\'en speed is 
maintained so that the magnetic field varies with density (B=$v_A$(4$\pi \rho$)$^{0.5}$). From the H$_2$ radial 
velocity structure described in \citet{2003RMxAA..39...77S} and  that the outflow H$_2$ knots have an average 
velocity projected onto the plane of the sky of $\sim$ 50~km~s$^{-1}$ (such a velocity is consistent with our 
modelling
results) we have estimated an inclination angle to the plane of the sky 
of between 5$^{\circ}$ and 10$^{\circ}$. The bows have been modelled propagating at this angle, 
i.e.~100$^{\circ}$ to the line of sight as the western outflow is redshifted. At this angle the model images closely
resemble the observations suggesting that the bow speed is similar to the shock speed, i.e the bows are propagationg in a
medium which is relatively at rest.   

The field angle, $\mu$, is the angle between the bow direction of motion and the magnetic field 
(see \citet{2004A&A...419..975O} for a
detailed description of the geometry). A misalignment of these directions 
results in asymmetric bow wings such as are observed for HH\,211. In our case the best results were found for $\mu$
$\sim$60$^{\circ}$$\pm$10$^{\circ}$. 

\begin{figure}
   \includegraphics[width=8.5cm]{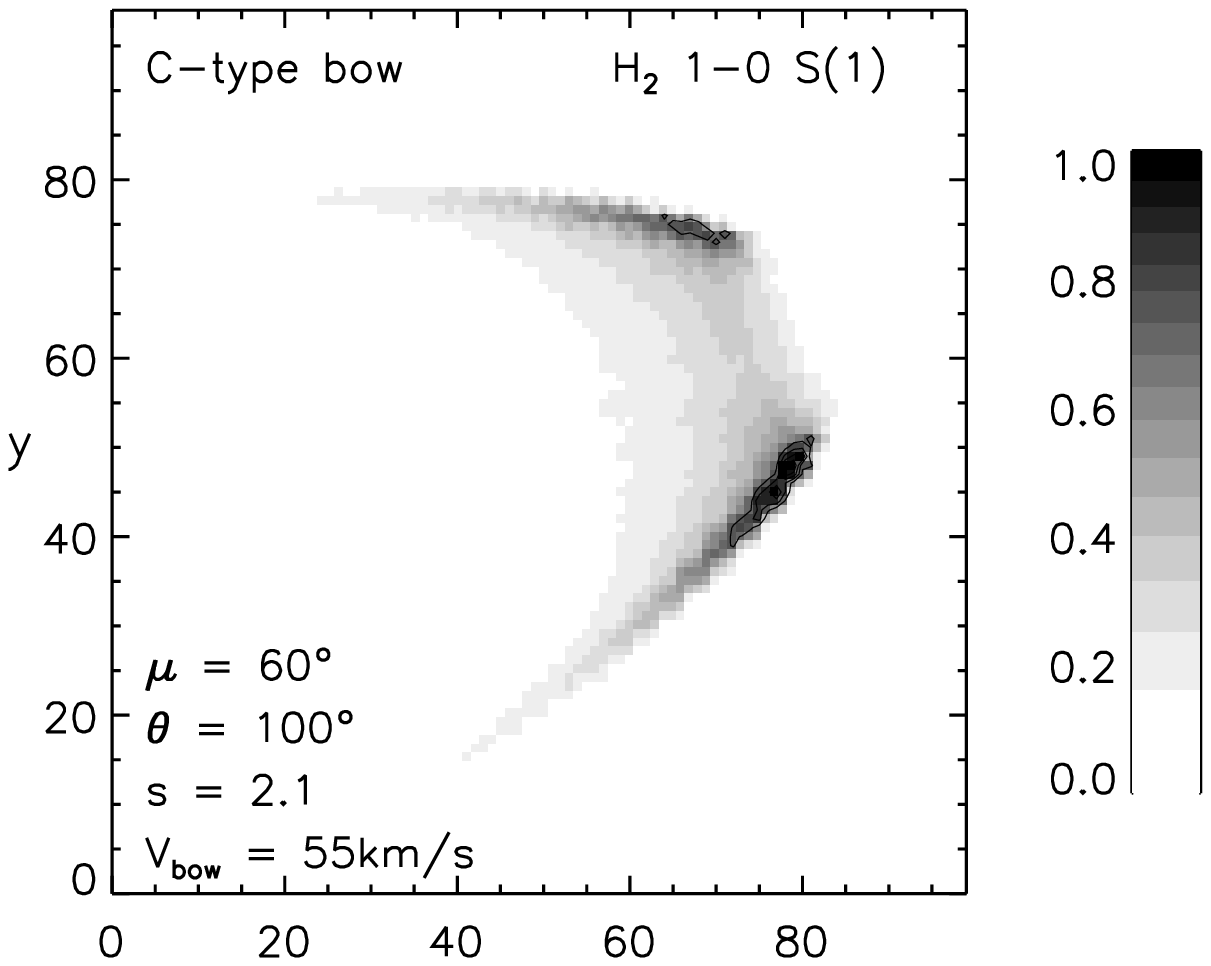}
   \includegraphics[width=8.5cm]{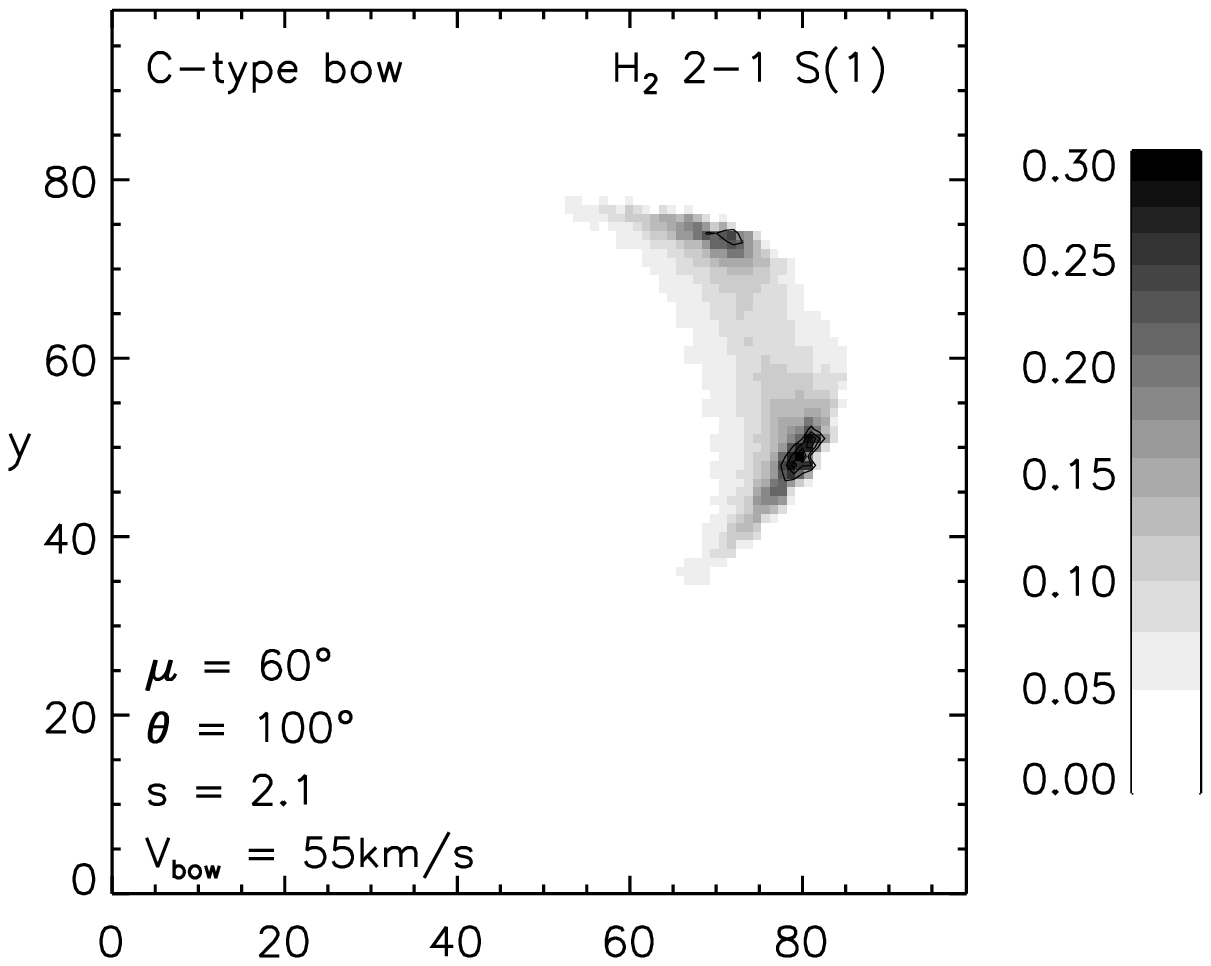}
   \includegraphics[width=8.5cm]{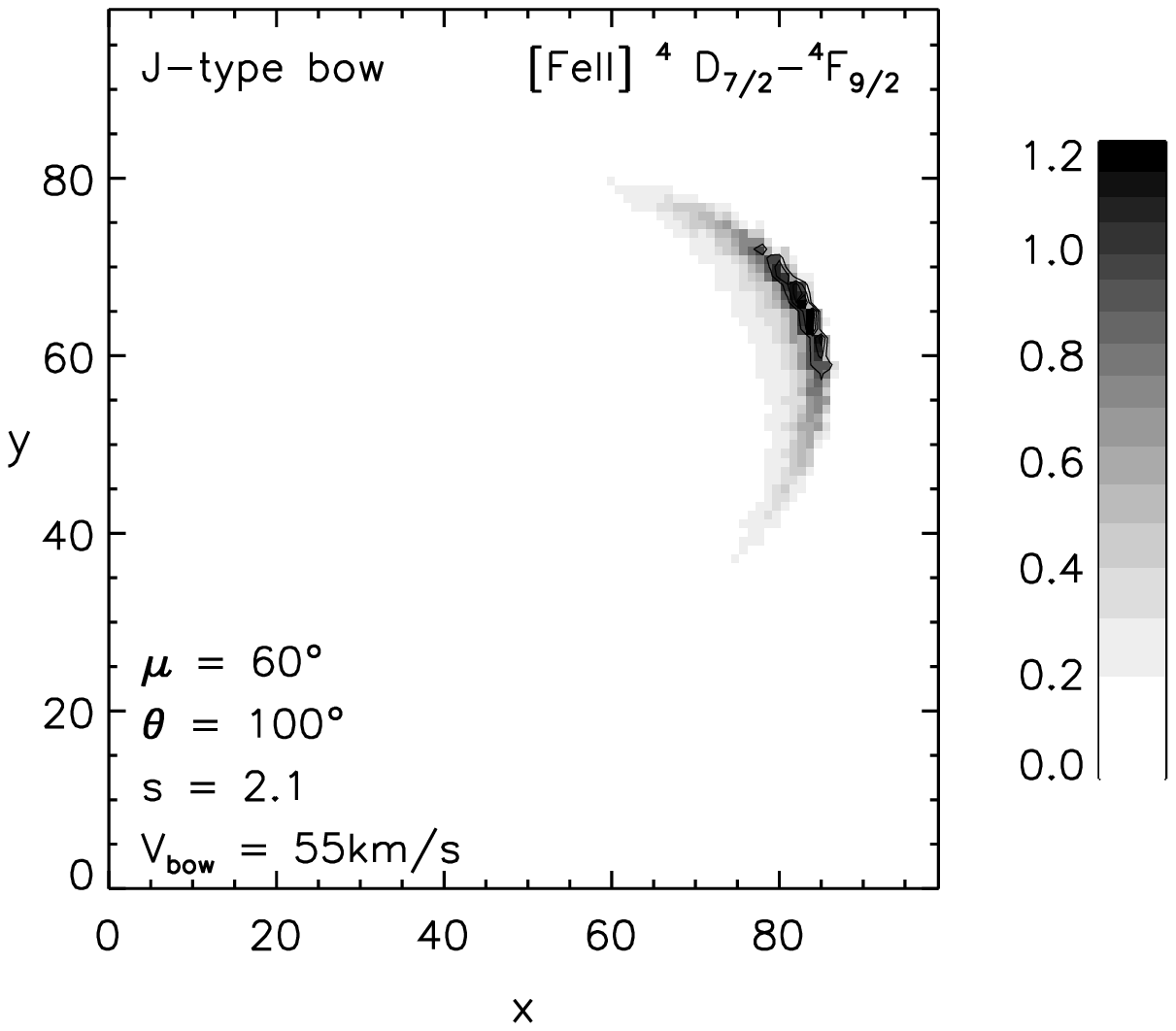}
      \caption{A C-type bow shock model for bow-{\em de} shown in H$_2$ (1,0)~S(1) and (2,1)~S(1) 
      excitation lines. Adopting a source distance
      of 315 pc gives a pixel scale along the x and y axes of 1 pixel = 0\arcsec .18. The bow direction 
      of motion 
      is inclined to the plane of the sky by an angle of 10$^{\circ}$, away from the observer, i.e. 
      100$^{\circ}$ to the angle of sight. The
      bow is moving at 55 km s$^{-1}$ relative to the ambient medium of H density,n, = 8 
      $\times$ 10$^3$ cm$^{-3}$. 
      All flux levels are 
      indicated normalised to the maximum H$_2$
      (1,0)~S(1) line flux. Hot dissociative shocks are necessary to induce emission from 
      [FeII]. Here we have employed a J-shock model (same parameters) to simulate the flux 
      distribution at 1.644 $\mu$m.}
\label{model_de}        
\end{figure}

Fig.~\ref{model_de} displays the simulated bow image to compare to bow-{\em de} in (1,0)~S(1) and (2,1)~S(1) 
ro-vibrational transition lines of H$_2$, as well as the [Fe{\small II}] 
${^4}$D$_{7/2}$ -- ${^4}$F$_{9/2}$ transition line. The shock speed is 55~km~s$^{-1}$ and the pre-shock density is 
8 $\times$ 10$^3$ cm$^{-3}$. The bow size, L$_{bow}$, has been chosen in order to match the distance 
between the upper and lower wings, in this case $\sim$5\arcsec. 

The [Fe{\small II}] emission is generated by a J-type dissociative bow and is 
restricted to a compact but elongated zone towards the bow apex where the highest temperatures are reached.
However, the observed [Fe{\small II}] 
emission is restricted to a single compact condensation, unlike the model distribution. According to this model the
total cooling in the (1,0)~S(1) line is 6.8$\times$10$^{-3}$ L$_{\sun}$, 
1.4$\times$10$^{-1}$ L$_{\sun}$ in all H$_2$ rotational and 
vibrational lines and 2.4$\times$10$^{-1}$ L$_{\sun}$ in all atomic and molecular lines.  

\begin{figure}
   \includegraphics[width=8.5cm]{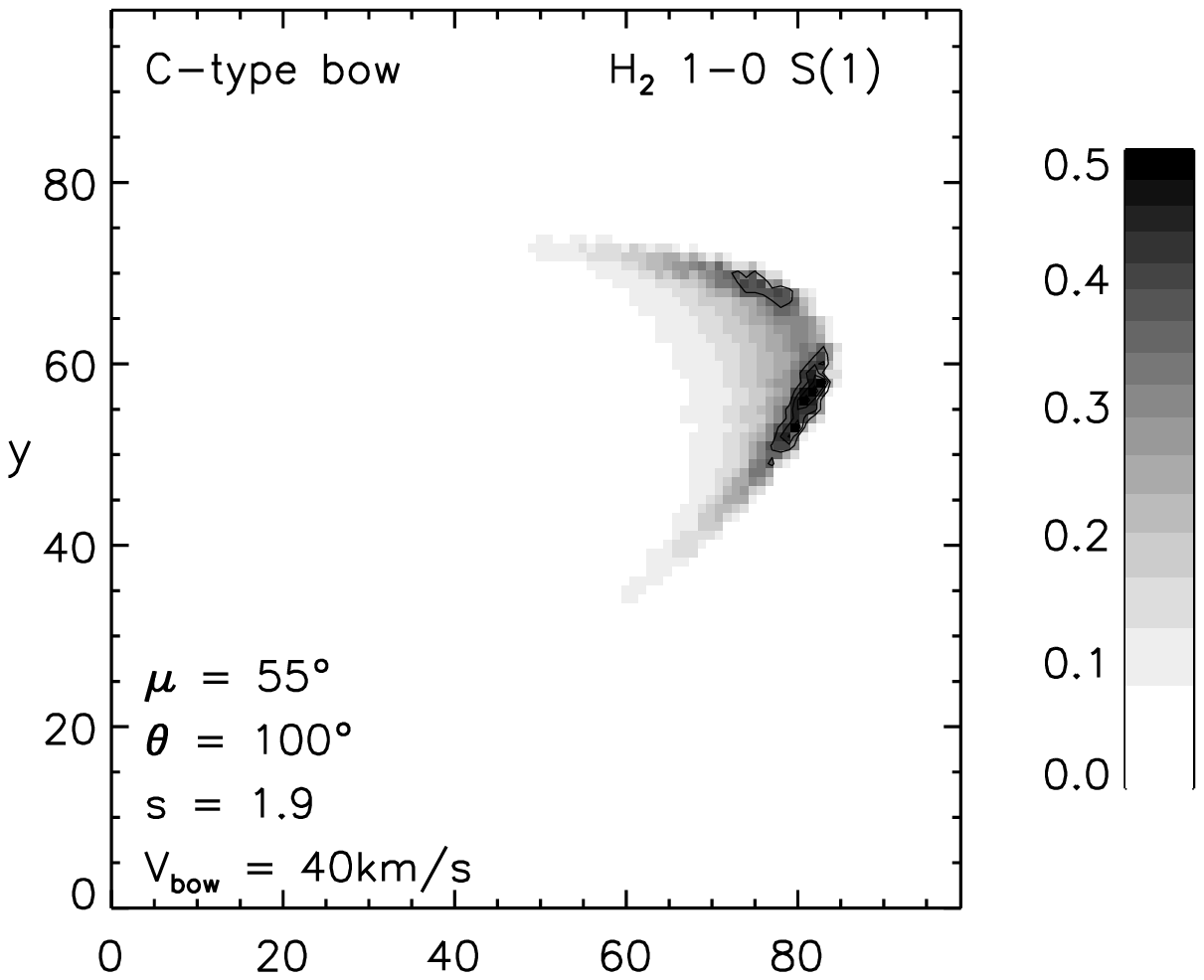}
   \includegraphics[width=8.5cm]{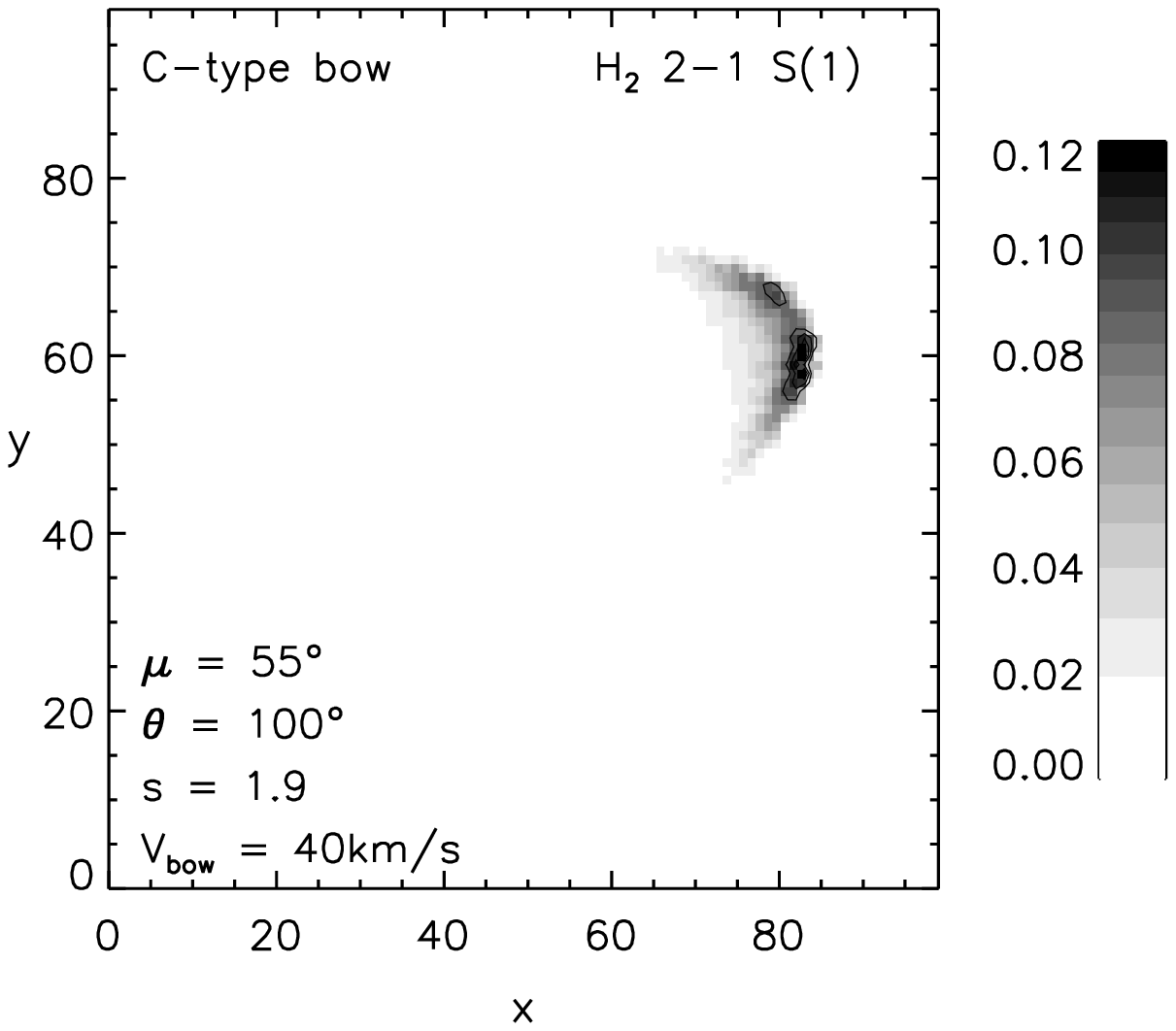}
        \caption{C-type bow shock model for bow-{\em bc} which reproduces the observed H$_2$ 
      (1,0)~S(1) and (2,1)~S(1) flux distribution. Adopting a source distance
      of 315 pc gives a pixel scale of 1 pixel = 0\arcsec .18. Here we take a lower bow speed of 
      40 km s$^{-1}$ and a pre-shock H density of 4 $\times$ 10$^3$ cm$^{-3}$ to match the 
      observed luminosity.
      Other parameters are given in table~\ref{cmodels}. The grey scale flux indications are 
      normalised to the maximum (1,0)~S(1) level for bow-de in Fig.~\ref{model_de}.}
\label{model_bc}         
\end{figure} 
   
Bow-{\em bc} is observed in both the (1,0)~S(1) and (2,1)~S(1) lines. Fig.~\ref{model_bc} displays the model 
generated images which closely resemble the observations. The distance between the upper and lower wings is
$\sim$3\arcsec. This knot is modeled with a reduced bow speed of 
40~km~s$^{-1}$ which is propagating into a lower density medium (n = 4.0 $\times$ 10$^{3}$ cm$^{-3}$. 
No [Fe{\small II}] emission is observed, consistent with the fact that its predicted luminosity lies 
below the detection threshold. According to this model the total cooling in 
the (1,0)~S(1) line is 1.4$\times$10$^{-3}$ L$_{\sun}$, 3.6$\times$10$^{-2}$ L$_{\sun}$  in all 
H$_2$ rotational and vibrational lines and 5.4$\times$10$^{-2}$ L$_{\sun}$ in all atomic and molecular lines.

Bow-{\em a} appears in the (1,0)~S(1) line as a compact knot of emission with a (2,1)~S(1) / (1,0)~S(1) ratio of 
0.08 $\pm$ 0.02. A bow speed of 29~km~s$^{-1}$ results in emission restricted to the bow apex. The calculated line 
cooling is 3.2\,$\times$\,10$^{-4}$ 
L$_{\sun}$ in the (1,0)~S(1) line, 8.8\,$\times$\,10$^{-3}$ L$_{\sun}$ in all H$_2$ lines and 
1.3\,$\times$\,10$^{-2}$ L$_{\sun}$ total line cooling.  
 \begin{figure}
   \includegraphics[width=8.5cm]{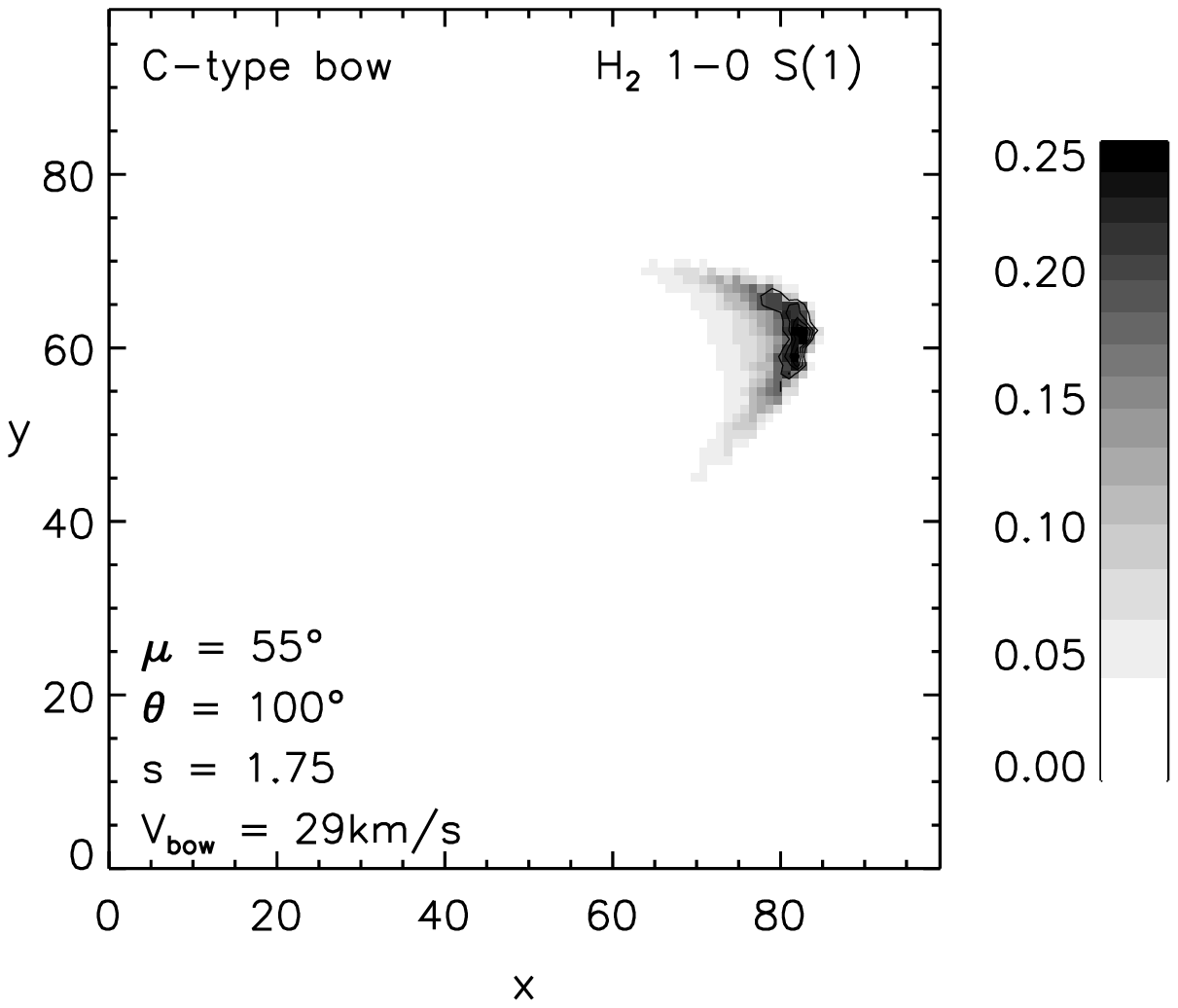}
       \caption{Faint (1,0)~S(1) emission is detected at bow-{\em a}. A bow propagating at 
       29 km s$^{-1}$ generates the correct (1,0)~S(1) luminosity with luminosities from 
       the other lines below the level of detectability, see table~\ref{lumtable}. Adopting 
       a source distance
       of 315 pc gives a pixel scale of 1 pixel = 0\arcsec .18. The grey scale 
       flux indications are 
       normalised to the maximum (1,0)~S(1) level for bow-de in Fig.~\ref{model_de}.}
 \label{model_a}          
   \end{figure}
   
We conclude that the observed structures are recreated in the model by solely altering the density, 
bow speed and ion fraction while all other parameters remain fixed. The extent of the bow is influenced
principally by the bow speed as it determines the location of the H$_2$ emission. A slower bow is 
characterised by emission concentrated closer to the bow front. For this reason we have maintained a 
constant L$_{bow}$ for all the models. The bow speed systematically decreases as an otherwise similar bow 
ploughs into a material of decreasing density. An increase in the ion fraction is expected in less dense 
regions where cosmic rays and UV photons can more easily penetrate the gas.
   
The driving power of a bow is converted into heat at a theoretical rate 
$P = \zeta \rho v_{bow}^3 L_{bow}^2 \pi / 2$ or
\begin{equation}
   P = 0.13 \zeta  \biggl( \frac{n}{8.0 \times 10^{3}~ {\rm cm}^{-3}}  \biggr)
       \biggl( \frac{v_{bow}}{55~ {\rm km~s}^{-1}} \biggr)^3 
       \biggl( \frac{L_{bow}}{1.0 \times 10^{16}~ {\rm cm}}\biggr)^2 {\rm L}_{\sun},
\label{eq:power}
\end{equation}
where $\zeta$ is a non-dimensional factor related to the aerodynamical drag and is of order unity. $\rho$ 
is the mass density equal to $2.32 \times 10^{-24}n$ ($n$ is the hydrogen nucleon density). 
The derived numbers are thus consistent with expectations.

\subsection{The outflow continuum emission and excitation}

\begin{figure*}
\hspace{-8mm}
   \includegraphics[width=19cm]{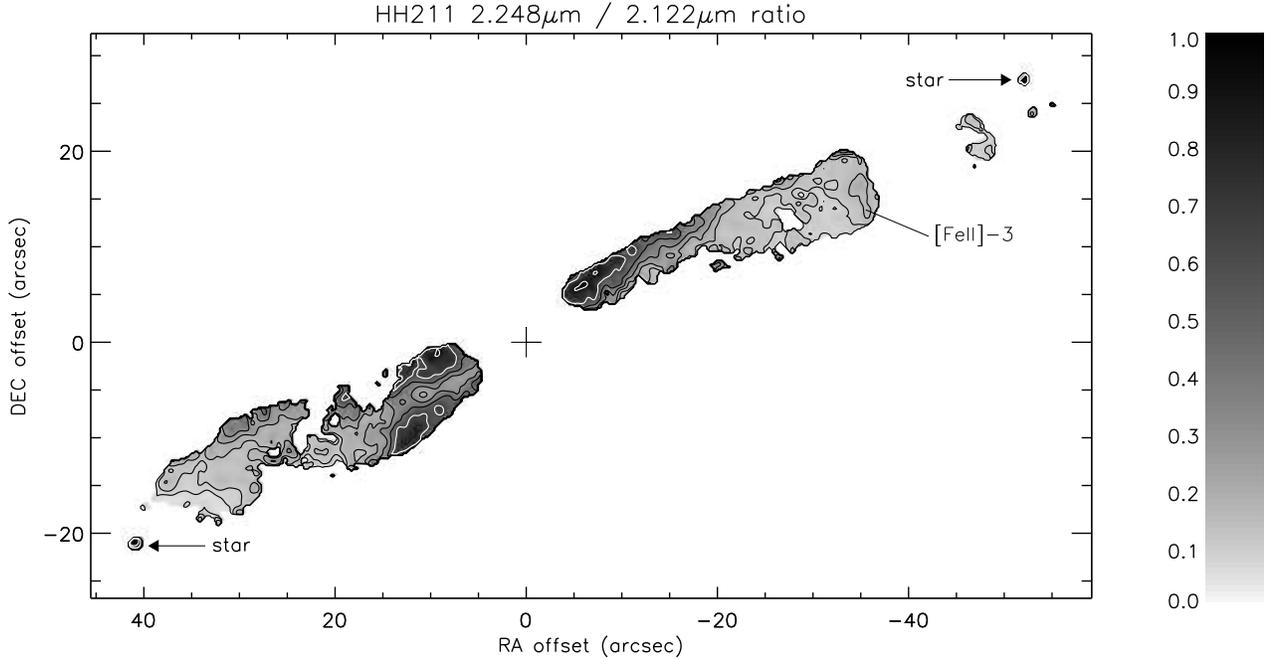}
      \caption{A map showing the ratio of our 2.122$\mu$m and 2.248$\mu$m images. The two images have not been
      continuum-subtracted (due to the very low S/N of the continuum image), so the map only reveals the H$_2$ (2,1)~S(1) \,/\,(1,0)~S(1) line ratio towards 
      the ends of the outflow, where the continuum is weak; the darker regions near the center of the image are where 
      the emission is dominated by the continuum. Logarithmically increasing 
      contour levels are at
      0.08, 0.11, 0.16, 0.22, 0.32, 0.45 (black) and 0.63, 0.89 (white).  }
\label{ratio_image}       
\end{figure*}
   
We have also divided the fluxes in the narrow-band images containing the (1,0)~S(1) flux and the (2,1)~S(1) flux 
to produce the ratio image displayed in Fig.~\ref{ratio_image}. In order to produce this map both images were 
first smoothed with a Gaussian FWHM = 0\arcsec.6 and values lying below the noise level were excluded before 
dividing. Note that the images were not 
continuum subtracted, due to the poor quality of the continuum image at 2.14$\mu$m, so Fig.~\ref{ratio_image} reveals two distinct outflow regions: (1) areas where the 
continuum emission is relatively strong possess a ratio above 0.3 and approach 1.0 where only continuum emission 
is present and (2) parts of the outflow not effected by continuum emission where the H$_2$ excitation ($\sim$
0.1) can be studied.

The outflow regions dominated by continuum emission are located along the edges of the SiO jet imaged by 
\citet{2001ApJ...555..139C}. This supports the idea that the continuum arises through scattered light from the 
protostar. The light escapes along a jet excavated cavity but not along the high density jet itself. It is 
cut off at 17\arcsec (8.0 $\times$ 10$^{16}$ cm) along the western outflow where it possibly encounters a high 
density 1 M$_{\sun}$ filament \citep{1999A&A...343..571G}. It is also possible that the filament lies in 
front of the flow. This projection effect would explain the higher extinction measured in the western outflow, 
contrary to the submillimeter dust emission maps of \citep{2000ApJ...530..851C} which do not reveal a strong 
asymmetrical density distribution between the eastern and western outflows. Continuum light encounters less 
hindrance  along the eastern outflow where it terminates alongside shock excited H$_2$ emission at 
knots {\em i} and {\em j}, 45\arcsec (2.1 $\times$ 10$^{17}$ cm) from its protostellar origin. 
Here continuum emission is seen at 2.14$\mu$m \citep{2003ApJ...595..259E} but not in our ratio image as the 
H$_2$ emission is relatively strong here. The origin and implications of the continuum emission will be 
discussed in \S~\ref{Discussion}.

The excitation ratio along the western outflow,$\sim$ 0.1, is typical of outflows seen in collisionally 
excited emission \citep{1976ApJ...203..132B,1982ARA&A..20..163S}. There is an 
increase in the excitation ratio towards the leading edge of bow-{\em de} where higher shock velocities and 
temperatures are reached and the H$_2$ line emission reaches its maximum value. Similar conditions were found
for the HH~240 bow shocks \citep{2004A&A...419..975O} and they are well explained through the bow shock
interpretation. However, a strong deviation from this picture is found in the [Fe {\small II}] image:
The localised [Fe{\small II}] emission is coincident with a region of higher ratio, (labelled [Fe{\small II}]--3) 
and, puzzlingly, not at the expected H$_2$ dissociated bow apex as seen in the model generated image, 
Fig.~\ref{model_de}. 
            
\section{Discussion}
\label{Discussion}

Which mechanisms give rise to the series of near-infrared bow shocks? Our
analysis suggests a combination of two processes: 

(1) The bow shocks are generated by a series of similar outflow accretion/ejection events. 
Using the model velocities the time lapses between the bows (and therefore between outflow events) are 
$\sim$~425 and $\sim$~290 years. These numbers are consistent with the detection of three bows given the 
dynamical age of $\sim$~1000 years. Fluctuations in jet activity probably manifest themselves as bow 
shocks at the jet cloud impact region which is Knot-{\em f} where the high speed CO jet terminates and 
H$_2$ shock heating is initialised. At Knot-{\em f} two H$_2$ (1,0)~S(1) velocity components have been 
observed by \citet{2003RMxAA..39...77S}. This reverse/transmitted shock pair seems to indicate the critical 
zone of jet impact where H$_2$ bow shocks are born. After formation, the bows propagate away from the
protostar towards the cloud edge and through a changing environment; they become less luminous as the density
decreases and they loose their momentum. Bow-{\em a} represents the final stages in the detectable life of one of these 
bow shocks. In support of this process, we note that the molecular jet demonstrates an apparent acceleration
along its length of value  $\sim$~5\,$\times$\,10$^{-3}$~km~s$^{-1}$~AU$^{-1}$
\citep{2001ApJ...555..139C}. Given ballistic motions, this implies 
that the entire jet now observed was ejected  within a relatively short period of time just 100--150 years ago.

(2) The bow shocks become illuminated within regions where the outflow
impacts on denser clumps of gas. This idea is supported by the implied high K-band extinctions of 2.9 and 1.8
magnitudes for knots {\em de} and {\em i}. These values do not represent the entire outflow and cannot be used to infer
the dereddened luminosity for the whole outflow. Additionally, strong continuum emission is detected in the
eastern outflow at knots {\em i} and {\em j} where the extinction is high (A$_K$ $\sim$ 1.8). If this is indeed 
scattered light from the protostar then it has channeled through to where the enhanced density has resulted in 
scattering. The passage of a C-shock will increase the post-shock density with a compression ratio of $\sqrt{2}$ 
times the magnetic Mach number (shock speed divided by the Alfv\'en speed) \citep{1978ppim.book.....S}. Therefore, the
continuum emission is likely to be seen alongside the shock excited H$_2$ emission as is the case for HH~211.      

\citet{1994ApJ...436L.189M} suggested an average A$_K$ of 1.2 magnitudes which is lower than the values 
that we derived in \S~3 but is consistent to within our error limits. Adopting this as the average 
extinction over the entire outflow we find that the total H$_2$ (1,0)~S(1) luminosity of the entire outflow is 
0.009~L$_{\sun}$. According to the predictions of our bow shock models, $\sim$\,4\% of the total H$_2$ line 
emission (2.7\% of total line cooling) is emitted in the (1,0)~S(1) which gives an intrinsic H$_2$ luminosity 
for HH~211 of $\sim$\,0.23~L$_{\sun}$ and a total luminosity, L$_{rad}$, resulting from all line emission of 
$\sim$\,0.34~L$_{\sun}$. If all the mechanical energy is converted into radiation then L$_{mech}$ is 
equivalent to L$_{rad}$ and L$_{mech}$ / L$_{bol}$ $\sim$ 10\%.   

Provided that the velocity of the swept up gas, which coincides with the CO outflow, roughly equals
the shock velocity (i.e. a radiative shock), \citet{1996A&A...305..694D} have shown that L$_{rad}$
should be roughly equal to the kinetic luminosity L$_{kin}$ of the outflowing material as measured 
through the CO luminosity. These criteria are indeed met for HH~211 as L$_{kin}$ $\sim$ 0.24 L$_{\sun}$ 
and the CO and H$_2$ are coincident suggesting that the shock is essentially radiative 
\citep{1999A&A...343..571G,2001ApJ...555...40G}. Note that the average extinction is restricted to about 
1 magnitude in the K-band in order to meet these criteria.

The driving source HH\,211--mm has a bolometric temperature of 33K which implies a youthful outflow 
system.
However, the HH~211 outflow itself does not reveal any characteristic signs of its assumed 
youthfulness besides the small 
spacial extent which implies a dynamical timescale of order 1000 years. This outflow age is limited 
by the density structure of the environment, as older bow shocks may simply have disappeared into 
sparse material. For this reason the outflow extent itself cannot be used to infer the full duration 
of outflow activity.

The high density jet (2--5 $\times$ 10$^6$ cm$^{-3}$) observable in SiO J = 5$\rightarrow$4 emission implies a
pre-shock density of about 2 $\times$ 10$^5$ cm$^{-3}$ \citep{2004ApJ...603..198G} and thus a maximum 
jet-to-ambient density ratio of $\sim$20. The CO J =  2$\rightarrow$1 maximum radial velocity is 
$\sim$~40~km~s$^{-1}$ \citep{1999A&A...343..571G} implying a jet velocity of 230\,--\,460~km~s$^{-1}$ 
given an inclination angle to the plane of the sky of between 5$^{\circ}$ and 10$^{\circ}$. Hydrodynamic 
numerical simulations find that a low density envelope generally develops around jets of density 
$\sim$~10$^{5}$cm$^{-3}$ and a jet-to-ambient density ratio of 10 \citep{1997A&A...318..595S,
1999A&A...343..953V,2004A&A...413..593R}. Could (pulsed) C-type jets which are heavier, denser and more 
ballistic than those simulated excavate  low density cocoons through which continuum 
emission might escape along the outflow? To date, such high density and high velocity jets involving C-type 
physics have not been simulated. The new challenge for theorists is to simulate such MHD jets.

HH~211 is of particular interest due to its unusually strong continuum emission. The original radiation may have
escaped from the source along a cavity of low optical depth which must have been excavated by powerful jet events. 
This radiation is then scattered when it encounters dense walls or clumps where the NIR optical depth along the
direction facing the source is of order unity. Some of the
scattered radiation then exits along the line of sight through a foreground of moderate NIR extinction.
This interpretation thus requires the original radiation to penetrate a distance of order 10$^{17}$\,cm with a mean density of
less than 10$^{5}$\,cm$^{-3}$ before encountering walls of thickness of order 10$^{16}$\,cm and density  
10$^{6}$\,cm$^{-3}$. A moderate fraction of the scattered radiation then escapes without encountering further dense
features along the line of sight. The high density and clumpiness of the jet as revealed through SiO observations 
\citep{2001ApJ...555..139C,2002A&A...395L..25N} are important factors to consider in  this interpretation.

The [Fe{\small II}] emission at 1.644 $\mu$m originates from an upper energy level of 11,300 K (compare to 6,953 K for
H$_2$ (1,0)\,S(1)) therefore we expect it to
highlight the hotter, high excitation regions of an outflow. As shown in \S~\ref{analysis}, the [Fe{\small II}]
emission from a typical bow shock should be located towards the front of the bow, where H$_2$ is dissociated. Clearly,
this is not the case for HH\,211 (bow--{\em de}). The most likely explanation is that the front of the bow shock is 
traversing a low density region. Material has been swept out by outflow activity and the bow shocks become luminous 
only where they interact 
with the wall of this hollowed out cavity. In this way it is possible see [Fe\,{\small II}]
emission coincident with bright H$_2$ emission in the `shoulders' of the bow shock. Such an explanation is also 
supported by the distinct possibility that continuum emission from the source is
escaping along an outflow excavated cavity.
  
\section{Conclusions}
\label{conclusions}

We have studied the HH\,211 protostellar outflow in the near-infrared regime through high resolution imaging
and spectroscopy. Images in the (1,0)~S(1) and (2,1)~S(1) ro-vibrational transitions of H$_2$ have been
analysed in order to study the excitation throughout the outflow. A narrow-band image at 1.644$\mu$m was
presented where the outflow is clearly seen and several confined condensations of emission are identified. In addition,
we have presented K-band spectroscopic fluxes for two separate prominent locations from which the extinction and excitation
conditions have been investigated. We have successfully modeled the series of bow-shocks in the western outflow 
as curved 3--dimensional shock fronts with steady state C--type physics. 

Our findings have lead to several conclusions about the nature of the outflow:

\begin{itemize}

\item C-type bow shocks propagate along the western flow. Model fitting has constrained several parameters 
including the density, bow velocity, ion fraction, intrinsic bow shape and magnetic field strength and direction.

\item The bow shocks are passing through dense clumps where their luminosity is accentuated. High values of
extinction  are measured in these regions. 

\item Bows {\em de}, {\em bc} and {\em a} are plausibly modeled as a series of initially identical 
bow shocks propagating through a medium of decreasing density. The bows slow down and disappear as
they approach the cloud edge. 

\item A misalignment of the magnetic field and outflow directions can account for the observed bow shock 
asymmetries. 

\item The most likely source of the protracted continuum emission is light from the protostar which evades dense 
core obscuration by escaping through a low density cavity excavated by the jet, as discussed in
\S~\ref{Discussion}. The continuum light is scattered 
when it encounters the denser material aligning the jet tunnel and the dense clumps along the outflow.   

\item The excitation along the outflow is typical of outflows in general. The ortho to para ratio of 3 
for molecular hydrogen implies shock heating as the source of the near-infrared line emission.

\item The [Fe{\small II}] emission is predicted and detected in isolated condensations. These condensations are
coincident with strong H$_2$ emission. However, the location of the [Fe{\small II}] emission is puzzling; 
it is not found in the expected bow apex region as predicted. This may also be due to the low density tunnel 
through which the bow apex is propagating.

\end{itemize}

These findings together with the large volume of previously published material is suggesting a 
global outflow model, as follows. Episodic fluctuations in accretion/ejection (of order a few hundred 
years) give rise to a variable jet velocity. The resulting shocks manifest themselves as C-type 
bow shocks at the principle jet/ambient medium impact region where they are detected outside the 
dense core where the extinction is lower. The bows propagate towards the cloud edge through a 
changing environment. The model suggests that the mean density decreases with distance from the core but 
that the bow shocks brighten where they encounter dense clumps. It is feasible that these clumps consist of
gas swept up by passage of previous bows driven by the alternating outflow power.

It is clear that in-depth studies of a wide range of protostellar outflows will yield valuable insight into how the
cloud environment sculpts the outflow and how much the environment itself has been influenced by the star forming
process.

\begin{acknowledgements}

This research is supported by a grant to the Armagh Observatory from the Northern Ireland Department of Culture,
Arts and Leisure and by the UK Particle and Astronomy Research Council (PARC). We would like to acknowledge the
data analysis facilities provided by the Starlink Project which is run by CCLRC / Rutherford Appleton laboratory 
on behalf of PPARC. This publication makes use of the Protostars Webpage hosted by the  Dublin Institute for Advanced
Studies.

\end{acknowledgements}

\bibliographystyle{bibtex/aa}
\bibliography{biblio}

\end{document}